\documentclass[aps,superscriptaddress,showkeys,pre,reprint]{revtex4-1}

\usepackage{amsmath}
\usepackage{amssymb}
\usepackage{graphicx}
\usepackage[none]{hyphenat}
\date{\today}\small
\usepackage{CJK}
\begin{document}
\begin{CJK*}{UTF8}{gbsn}
\title{Modeling Epidermis Homeostasis and Psoriasis Pathogenesis}

\author{Hong Zhang (张红)}\thanks{These authors contributed equally}
\affiliation{CAS--MPG Partner Institute and Key Laboratory for Computational 
Biology, Shanghai Institutes for Biological Sciences, Shanghai 200031, China}
\affiliation{Naval Submarine Academy, Qingdao, Shandong 266000, China}
\author{Wenhong Hou (侯文洪)}\thanks{These authors contributed equally}
\affiliation{CAS--MPG Partner Institute and Key Laboratory for Computational 
Biology, Shanghai Institutes for Biological Sciences, Shanghai 200031, China} 
\author{Laurence Henrot}
\affiliation{Sprim Advanced Life Sciences, 1 Daniel Burnham Court, San Francisco, CA 94109, USA}
\author{Marc Dumas}
\author{Sylvianne Schnebert}
\author{Catherine Heus}
\affiliation{LVMH Research, 185 Avenue de Verdun, 45804, Saint-Jean-de-Braye, France}
\author{Jin Yang (杨劲)}\thanks{Correspondence. 320 Yue Yang Road, Shanghai  200031, 
China. Tel: +86-21-54920476; Fax: +86-21-54920451; E-Mail: 
jinyang2004@gmail.com}
\affiliation{CAS--MPG Partner Institute and Key Laboratory for Computational Biology, Shanghai Institutes for Biological Sciences, Shanghai 200031, China}

\begin{abstract}
We present a computational model to study the spatiotemporal dynamics of the epidermis homeostasis under normal and pathological conditions. The model consists of a population kinetics model of the central transition pathway of keratinocyte proliferation, differentiation and loss and an agent-based model that propagates cell movements and generates the stratified epidermis. The model recapitulates observed homeostatic cell density distribution, the epidermal turnover time and the multilayered tissue 
structure. We extend the model to study the onset, recurrence and phototherapy-induced remission of
psoriasis. The model considers the psoriasis as a parallel homeostasis of normal and psoriatic keratinocytes originated from a shared stem-cell niche environment and predicts two homeostatic modes of the psoriasis: a disease mode and a quiescent mode. Interconversion between the two modes can be controlled by interactions between psoriatic stem cells and the immune system and by the normal and psoriatic stem cells competing for growth niches. The prediction of a quiescent state potentially explains the efficacy of the multi-episode UVB irradiation therapy and recurrence of psoriasis plaques, which can further guide designs of therapeutics that specifically target the immune system and/or the keratinocytes.
\end{abstract}

\keywords{Mathematical model, Epidermal homeostasis, Psoriasis, Bimodal switch, Immune system}
\maketitle
\end{CJK*}
\small

\section{Introduction}

The epidermis, the outermost layer of skin, provides the human body a 
physiological barrier to the environment and protects the body from water loss, 
pathogenic infection and physical injury. The epidermis organizes into a 
stratified structure of keratinocytes at several differentiated 
stages~\cite{montagna1992atlas}, which constitute 95$\%$ cell population in the 
epidermis~\cite{rook2010rook}. Like other regenerative tissues, the epidermis 
constantly renews itself to replace desquamated and apoptotic keratinocytes, 
repair tissue damage and establish the homeostasis. The renewal is orchestrated 
by a cascade of cellular processes including proliferation, differentiation, 
migration, apoptosis and 
desquamation~\cite{weinstein1984cell,loeffler1987epidermal, 
milstone2004epidermal}. A keratinocyte transits spatially from the stratum 
basale to the stratum corneum during its lifespan and meanwhile experiences 
multi-stage biochemical and morphological changes. Many endogenous and exogenous 
factors (e.g., Ca$^{2+}$ concentration, cytokines, UV irradiation, etc.) affect 
the epidermal dynamics and the homeostasis of the epidermis by regulating one or 
more cellular processes.

Mathematical and computational models have long been useful tools to predict 
cellular behaviors of the epidermis renewal under normal or pathological 
conditions. Previous models for the epidermal dynamics usually adopted two 
approaches. One approach includes deterministic models that derived analytical 
solutions to stationary cell populations. For example, the model by 
Savill~\cite{savill2003mathematical} described proliferation of stem cells and 
transit-amplifying cells and differentiation to post-mitotic cells, which 
predicted the influences of apoptosis, cell-cycle time and transit time on cell 
populations. Gandolfi et al.~\cite{gandolfi2011age} proposed a spatiotemporal 
model to investigate the evolution of epidermis, which described cell motion by 
a constitutive equation. The other approach includes agent-based models that 
treat individual keratinocytes as computing entities operating under specific 
physical and biological rules. Such models can simulate the multi-layer 
epidermal structure organized by cell proliferation, differentiation, death and 
migration,  in which nonspecific intracellular and extracellular biochemical 
factors affected cell proliferation and differentiation while physical adhesive 
and repulsive forces governed cell 
motion~\cite{stekel1995computer,grabe2005multicellular,sutterlin2009modeling,
sun2007integrated,adra2010development}.

\begin{figure*}[t]
\centering
\includegraphics[scale=0.45]{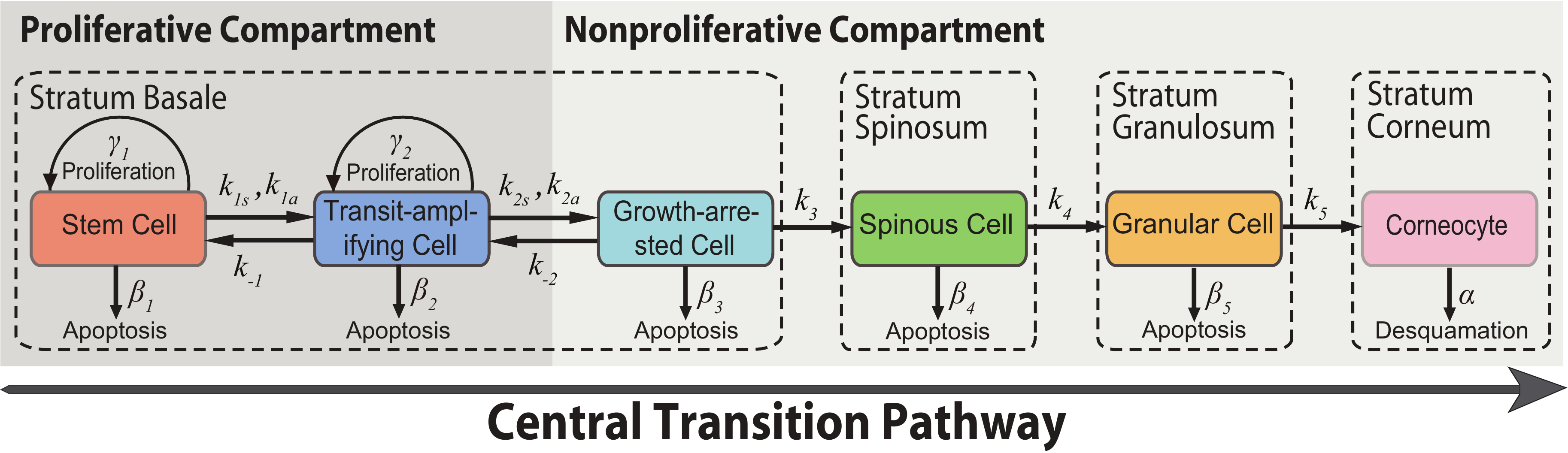}
\caption{{\bf The central transition pathway of the epidermis renewal.} The 
model describes proliferating and nonproliferating keratinocytes. Stem cells and 
transit amplifying cells proliferate by self-renewal and by symmetric division 
and asymmetric division. The transit-amplifying cells differentiate into the 
growth-arrested cells in the nonproliferative compartment, which in turn 
differentiate into keratinocytes of later stages, including spinous cells, 
granular cells and corneocytes. Nucleated cells undergo apoptosis and 
corneocytes desquamate from the stratum corneum. All cellular processes are 
parameterized by rate constants labeled on transitions.}\label{fig:model}
\end{figure*}

In this paper, we present a hybrid model to combine advantages of the above two 
approaches. The model uses a mean-field cell population kinetics together with 
an agent-based model for cell migration. The model computes population dynamics 
of the epidermal renewal without having to compute the cell movements 
simultaneously, allowing fast and analytical evaluations of modeling hypotheses 
and results. The population kinetics model describes cellular processes 
including cell division, differentiation, apoptosis and desquamation. Either 
deterministic or stochastic simulation can be used to generate the population 
dynamics of keratinocytes. The cell migration is described by a two-dimensional 
agent-based model that tracks the cell movement driven by cell-cell 
interactions. Simulation of the cell migration can be integrated with the 
stochastic population dynamics to visualize the tissue stratification and 
establishment of homeostasis. A properly parameterized model reproduces 
experimentally-observed epidermis growth, differentiation and desquamation 
dynamics, homeostatic density distribution over different types of keratinocytes 
and the epidermis turnover times of different cell compartments.

To investigate an important pathological condition of the skin, we study the 
onset and recurrence of psoriasis plaques and their recovery under the 
phototherapy by UVB irradiation. The psoriasis is a complex epidermal disorder 
characterized by keratinocyte hyperproliferation and abnormal differentiation 
due to intricate interactions with the immune system. The disease affects 2-4\% 
of the general population and currently has no 
cure~\cite{parisi2013global,meglio2014}. Specifically, we hypothesize a novel 
mechanism of stem cell-immune system interaction and predict the chronic 
disorder as a bimodal switch between a disease phenotypic and a quiescent 
(seemingly-normal) state. The model hypothesizes a parallel epidermal 
homeostasis simultaneously maintained by the normal and the psoriatic 
keratinocytes. The psoriatic homeostasis is caused by permanent perturbations in 
cell division, apoptosis and differentiation, derived from defective stem cells 
and their interactions with a weakened immune system. For treatment, the model 
predicts that to achieve a controlled remission the effective treatment of UVB 
irradiation must reduce the high-density psoriatic epidermis below a threshold 
level by activating the apoptosis in stem cells, potentially explaining the 
chronicity and recurrences of the disorder and providing a guide to design 
feasible therapeutics.

\section{The model}

The model consists of (1) a kinetics model, which tracks the temporal evolution 
of cell population of keratinocytes at several differentiation stages,  and (2) 
a migration model, which describes motion of individual cells to generate the 
stratified structure of the epidermis. One can choose the kinetic model as a 
standalone module to compute the mean-field population dynamics or can 
integrate the two models to visualize the renewal kinetics and stratification 
of the tissue. 

\subsection{Model of cell population kinetics}

Figure~\ref{fig:model} illustrates the {\em central transition pathway} of the 
epidermis renewal. The pathway considers the population dynamics of six 
categories of keratinocytes stratified from the stratum basale to the stratum 
corneum, including progenitors, stem cells (SC) and transit-amplifying (TA) 
cells in the proliferative compartment, and differentiated cells in the 
nonproliferative compartment with growth-arrested (GA) cells, spinous (SP) 
cells, granular cells (GC) and corneocytes (CC). The above classification is 
primarily based on known histological structure and physiological function of 
the human epidermis~\cite{denecker2008caspase}. Growth-arrested cells are 
precursors for nonproliferating cells. Spinous cells and granular cells are 
fully differentiated keratinocytes, and the nonnucleated corneocytes represent 
the end-stage differentiation and eventually desquamate. The central transition 
pathway incorporates three main cellular processes: (1) proliferation of stem 
cells and transit-amplifying cells, (2) differentiation including several 
inter-category cell conversions,  and (3) cell loss including apoptosis of 
nucleated cells and desquamation of the corneocytes. 

\begin{figure}[b]
\centering
\includegraphics[scale=0.35]{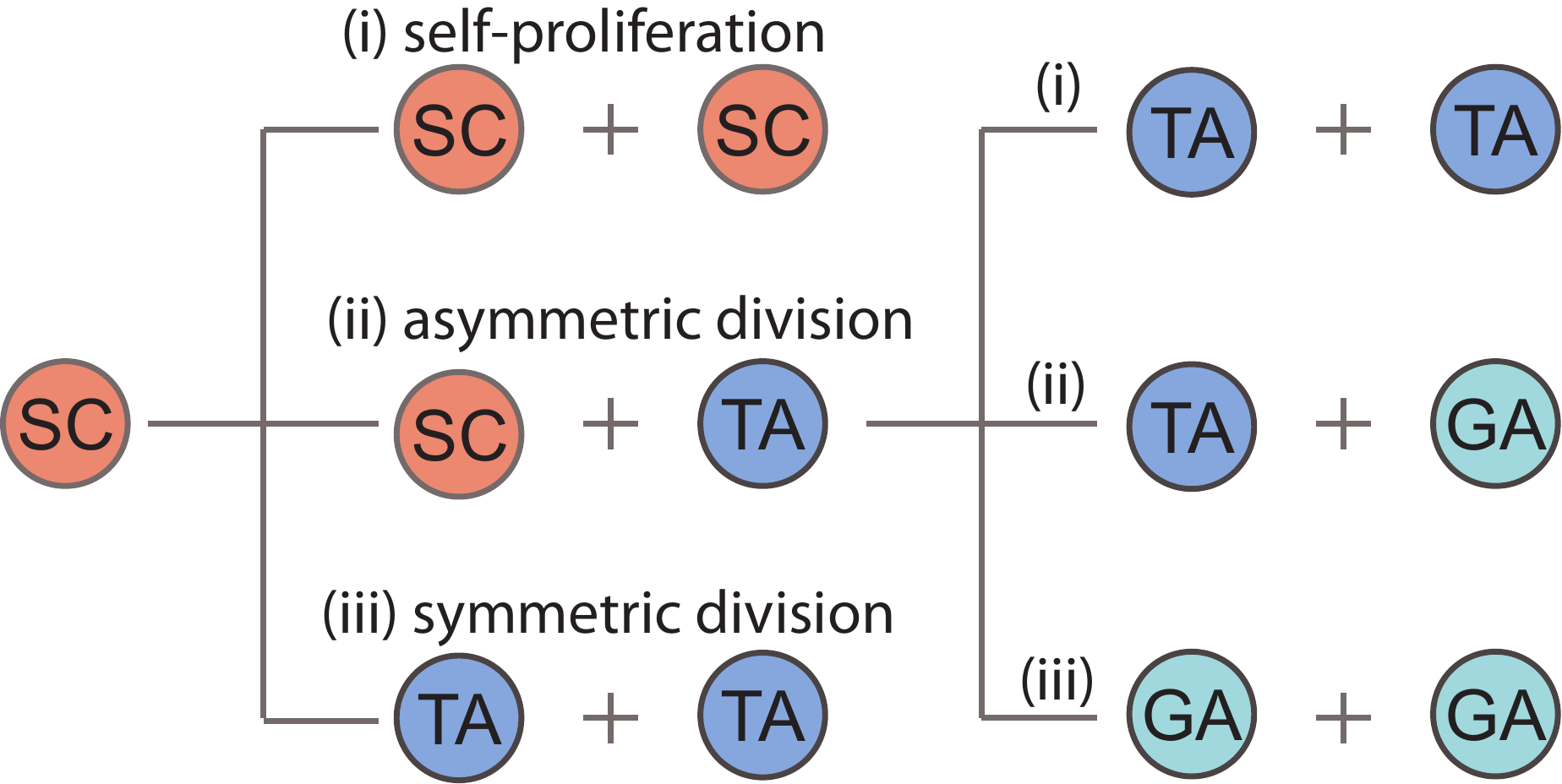}
\caption{\label{fig:division} {\bf Proliferation of stem cell and TA cell.} Both 
stem cell and TA cell undergo three types of divisions: (i) self-proliferation, 
(ii) asymmetric division and (iii) symmetric division. } 
\end{figure}

How stem cells maintain the epidermal homeostasis remains 
unresolved~\cite{de2012cell}. However, recent long-term {\it in vivo} lineage 
tracing studies on mouse tail tissue by genetic labeling revealed remarkable 
details about clonal dynamics of epidermal progenitors, suggesting the 
existence of either single (stem cell alone) or two progenitors (stem cell and 
committed 
progenitor)~\cite{clayton2007single,doupe2010ordered,mascre2012distinct}. As 
illustrated in Fig.~\ref{fig:division}, our model considers a slow cycling stem 
cell population together with a faster proliferating committed progenitors (or, 
transit-amplifying cells). A stem cell divides in one of three 
modes~\cite{watt2000out,simons2011strategies,simons2011stem}: 
(i) self-proliferation, by which a stem cell divides into two daughter stem 
cells, (ii) asymmetric division, by which a stem cell divides into a stem cell 
and a TA cell, or (iii) symmetric division, by which a stem cell divides into 
two TA cells. Considering a finite availability of stem  
cell niches~\cite{watt2000out,tumbar2004defining}, we assume a logistic growth 
of stem cells to limit the stem cell density by a maximal growth capacity, 
which ensures the system to reach a well-defined steady state (see 
Ref.~\cite{warren2009cells} for a more general model that guarantees a steady 
state). Previous models~\cite{clayton2007single,klein2007kinetics} required a 
precise balance between stem cell self-proliferation and symmetric division and 
were intolerable to arbitrary perturbations such as population random drift 
caused by intrinsic stochasticity in the three-mode stem cell division. 
Similarly, TA cells also divide in one of the three modes of 
self-proliferation, symmetric and asymmetric division into the GA 
cells~\cite{clayton2007single,Roshan2012884,alcolea2013tracking}. In addition, 
a TA cell may resume the stem cell state and a GA cell may resume a TA cell 
state by backconversions~\cite{okuyama2004dynamic}.  

The rate of progenitor division is often characterized by the cell-cycle time 
and by the subpopulation of cells that are active to divide (also known as the 
``growth fraction"). Our mean-field model does not distinguish proliferative 
propensity in individual cells and therefore parameterizes cell divisions with 
empirical rate constants that integrate influences of the cell cycle and the 
growth fraction. Environmental changes regulate the proliferation rate of stem 
cells. For example, the need of repairing tissue damage promotes stem cell 
proliferation~\cite{mascre2012distinct,simons2011strategies,
blanpain2009epidermal}. Recent study of hair follicles showed that TA cells may 
signal to stem cells to regulate proliferation~\cite{hsu2014transit}. To 
incorporate this feedback mechanism, we assume an empirical dependence of stem 
cell division rate constants, $\gamma_1$, $k_{1a}$ and $k_{1s}$, on the density 
of TA cells and define:
\begin{equation}\label{eq:scrate}
\frac{\gamma_1}{\gamma_{1,h}}=\frac{k_{1a}}{k_{1a,h}}=\frac{k_{1s}}{k_{1s,h}}
=\frac{\omega}{1+(\omega-1)(p_{ta}/p_{ta,h})^n} \ ,
\end{equation}
where the subscript $h$ indicates a homeostatic rate constant (see 
Table~\ref{tab:parameter} for numerical values), and $\omega\equiv r_{x,\rm 
max}/r_{x,h}$ is the ratio of the maximum division rate to the homeostatic rate 
and is assumed identical for all division processes. $\omega$ reflects the 
maximum increase in the growth fraction and/or decrease in the cell cycle time 
when stem cell proliferation accelerates. At homeostasis of the normal 
epidermis, the reported growth fraction varied from 20\% to 
70\%~~\cite{heenen1998ki,heenen1997growth}. Study in mice epidermis found more 
than 10-fold decrease in the cell cycle time from 5-7 days to 11 hours following 
tissue abrasion~\cite{morris1983epidermal}, whereas no significant change in 
cell cycle time was found in 
psoriasis~\cite{castelijns1998epidermal,castelijns2000cell}. The exponent $n$ 
models the sensitivity to the deviation of TA density from the homeostasis. In 
the limit of a much reduced TA cell density ($p_{ta}\ll p_{ta,h}$), stem cells 
divide at the maximum rate, $r_{x,\rm max}$, whereas stem cells divide at a 
minimal rate when TA cells overpopulate ($p_{ta}\gg p_{ta,h}$). The total stem 
cell proliferating rate at homeostasis is set about 10$^{-2}$ per day, aligned 
with 4-6 division events per year~\cite{mascre2012distinct}.

\begin{figure}[!t]
\begin{center}
\includegraphics[scale=0.28]{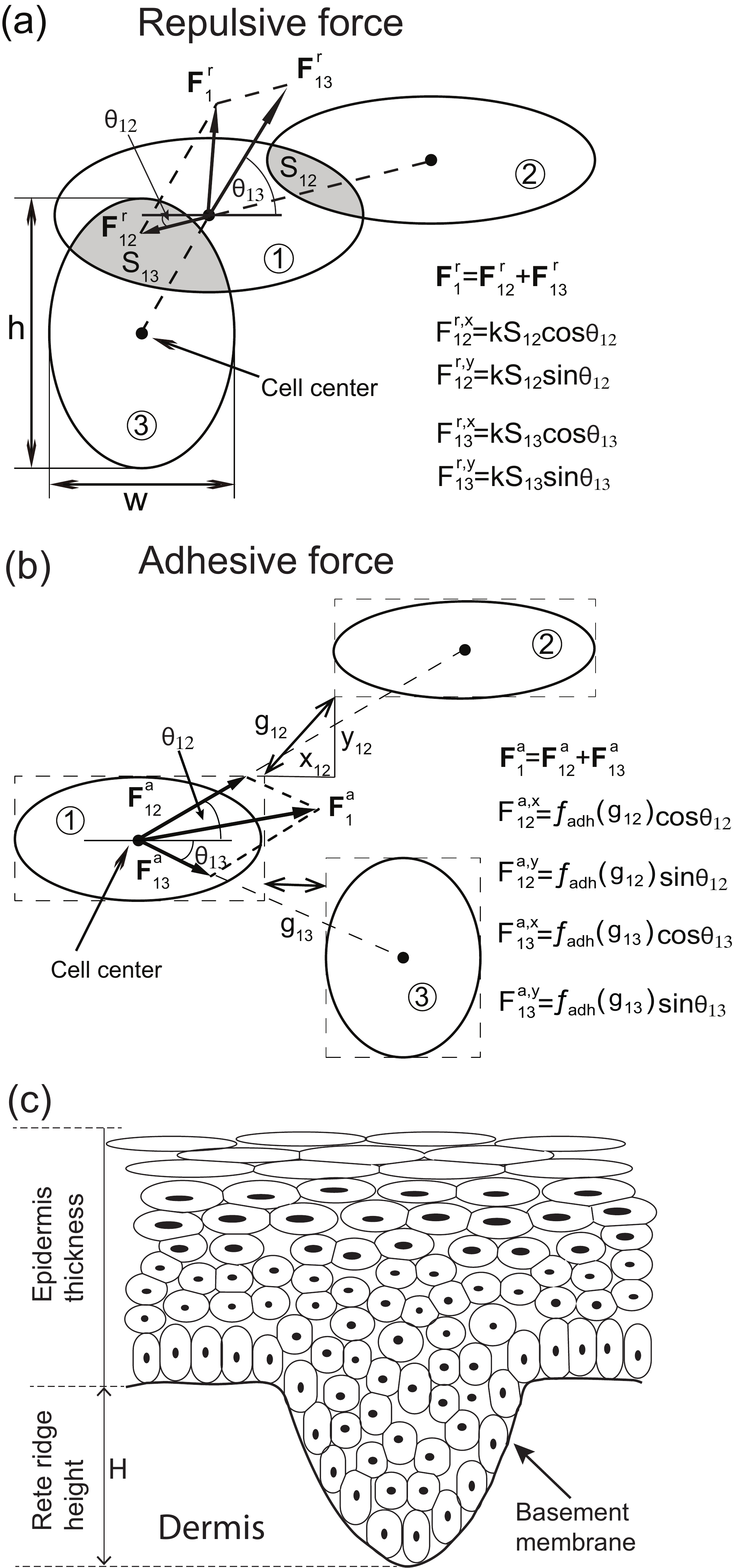}
\end{center}
\caption{ {\bf Mechanics of cell migration.} Keratinocytes are geometrically modeled as irrotational ellipsoids with two principal axes staying horizontal or vertical. (a) A repulsive force is determined by an overlap between neighboring cells. The repulsion ${\bf F}_{12}^r$ between cells 1 and 2 is proportional to the overlapping area $S_{12}$ approximated by the corresponding overlap rectangle. ${\bf F}_{13}^r$ is determined similarly. The net repulsion on cell 1 is a vector sum ${\bf F}_1^r={\bf F}_{12}^r+{\bf F}_{13}^r$. (b) The adhesive force exists between two adjacent cells. The force ${\bf F}_{12}^a$ is related to the distance between cell 1 and cell 2. The direction of ${\bf F}_{12}^a$ or ${\bf F}_{12}^r$  acting on cell 1 is paralleled to the line (dashed) connecting centers of cell 1 and cell 2. The repulsive force and adhesive force between two overlapped cells are balanced by the force generated by viscosity due to cell motion. Cell motion is only translational without rotation. $w$ and $h$ denote the width and height of a cell. (c) Epidermis thickness, rete ridge height, and undulant basement membrane geometrically configure the epidermis in a 2D projection.}\label{fig:force}
\end{figure}

The nonproliferative compartment describes a cascade of differentiations from GA 
cells to corneocytes. The model also considers an apoptosis process for all 
nucleated keratinocytes. Early studies suggested that apoptosis was only 
significant for stem cells and transit-amplifying cells in the proliferative 
compartment~\cite{laporte2000apoptosis}. Recent 
experiment~\cite{gagna2009localization} also showed that apoptosis was evident 
in the differentiated keratinocytes. The extent of apoptosis is commonly 
characterized by {\em apoptotic index}, which is typically quantified using the 
TUNEL (transferase-mediated uridine nick end labelling) 
assay~\cite{laporte2000apoptosis}. For a cell type $i$, the apoptosis index 
equals the probability that a cell undergoes apoptosis, defined as the ratio of 
apoptosis rate $f_i^{\rm apop}$ to the total outflux $f_i^{\rm out}$ including 
rates of apoptosis and transition to the downstream cell category:
\begin{equation}{\label{eq:apoptosis}}
{\rm AI}=\frac{f_{i}^{\rm apop}}{f_i^{\rm out}}=\frac{\beta_i}{\beta_i+k_i},  \ 
\ i=1,...5 \ ,
\end{equation}
For stem cells and TA cells, $k_1=k_{1s}$ and $k_2=k_{2s}$ are symmetric 
division rate constants. Experiments often reported global apoptosis indices 
that did not differentiate apoptotic activities in proliferative and 
non-proliferative compartments. Apoptotic heterogeneities of cells within a 
compartment is even less known. Here each cell category is assumed an identical 
apoptosis index that is used to calculate the apoptosis rate constants 
$\beta_i's$. Governing system equations (ODEs) are listed in 
Table~\ref{tab:equations} and computation of the model is described in the 
Supplementary Material (Fig.~S1).

\begin{table*}[t]
\small
\caption{ \bf Kinetic equations for the epidermal renewal and homeostasis}
\begin{tabular}{ll}\hline
 {\bf Cell type} & {\bf Rate equation} \\\hline\hline
Stem cell (SC) &${dp_{sc}/dt=[\gamma_1(1-p_{sc}/p_{sc}^{\rm max})-k_{1s}-\beta_1]p_{sc}+k_{-1}p_{ta}}$ \\
Transit-amplifying (TA) cell & ${dp_{ta}/dt=(\gamma_2-k_{2s}-\beta_2-k_{-1})p_{ta}+(k_{1a}+2k_{1s})p_{sc}+k_{-2}p_{ga}}$\\
Growth-arrested (GA) cell  & ${dp_{ga}/dt=(k_{2a}+2k_{2s})p_{ta}-(k_{-2}+k_3+\beta_{3})p_{ga}}$\\
Spinous cell (SP)& ${dp_{sp}/dt=k_3p_{ga}-(k_4+\beta_{4})p_{sp}}$\\
Granular cell (GC) & ${dp_{gc}/dt=k_4p_{sp}-(k_5+\beta_{5})p_{gc}}$\\
Corneocytes (CC)& ${dp_{cc}/dt=k_5p_{gc}-\alpha p_{cc}}$\\\hline
\multicolumn{2}{l}{\footnotesize $p_{sc},p_{ta},p_{ga},p_{sp},p_{gc}$, and $p_{cc}$ are cell densities.}
\end{tabular}\label{tab:equations}
\end{table*}

\subsection{Model of cell migration}

The agent-based migration model describes movement of all keratinocytes in a 
two-dimensional (cross-sectional) epidermis volume. Keratinocytes once derived 
from stem cells move outward from the stratum basale to the outermost stratum 
corneum, to compose a stratified epidermis. The model describes cell mechanics 
that propels cell movement. An individual cell is subject to three forces: (i) 
a viscous force due to cell moving in the surrounding environment; (ii) a 
repulsive force due to cell-cell compression; and (iii) an adhesive force due 
to interactions among adhesive molecules on cell membranes. Considering the 
sluggish keratinocyte motion (in a scale of $\mu$m/hr) in a fluidic environment 
with a low Reynolds number (i.e., viscosity dominates 
inertia)~\cite{purcell1977life}, the model neglects the acceleration due to 
inertia. The model also considers keratinocytes as non-chemical tactic cells 
that do not move by self-propulsion. The force balance for the $i$th cell is 
\begin{equation}\label{eq:force}
\mu\frac{d{\bf x}_i}{dt}+{\bf F}_i^r+{\bf F}_i^a=0 \ ,
\end{equation}
where vector ${\bf x}_i$ is the cell-center coordinate, and $\mu$ is the 
viscosity coefficient. The first term in Eq.~(\ref{eq:force}) is the viscosity 
of the epidermis. ${\bf F}_i^r$ and ${\bf F}_i^a$ are repulsive and adhesive 
forces between neighboring cells, which are sums of forces derived from all 
individual pairwise contacts,
\begin{equation}\label{eq:force1}
{\bf F}_i^r=\sum_{j=\mathcal{O}(i)}{\bf F}_{ij}^r \ , \ \ \ {\bf 
F}_i^a=\sum_{j=\Omega(i)}{\bf F}_{ij}^a \ ,
\end{equation}
where $\mathcal{O}(i)$ and $\Omega(i)$ denote sets of cells overlapping and 
neighboring with the $i$th cell and ${\bf F}_{ij}^r$ and ${\bf F}_{ij}^a$ are 
force vectors produced onto cell $i$ by interaction (repulsion or adhesion) 
between cells $i$ and $j$. By symmetry, ${\bf F}_{ij}^r=-{\bf F}_{ji}^r$ and 
${\bf F}_{ij}^a=-{\bf F}_{ji}^a$. The model treats individual cells as 
rigid-body agents and uses the extent of virtual cell overlap to determine the 
repulsive force [Fig.~\ref{fig:force}(a)]. Adhesion between two cells is a 
function of their spatial distance~\cite{palsson2008,li2013skin} 
[Fig.~\ref{fig:force}(b)]. Computation of ${\bf F}_{ij}^r$ and ${\bf F}_{ij}^a$ 
is given in the Supplementary Material.

Previous agent-based 
models~\cite{grabe2005multicellular,sun2007integrated,schaller2007modelling} 
treated keratinocytes as identically-sized circles or spheres. However, cells 
progressively adopt varied shapes and sizes at different stages of 
differentiation. Cells in an outer layer generally have more flattened cell body 
and larger surface area, compared to cells in layers underneath. Tissue location 
in the body can also influence the cell geometry. For example, compared to the 
mean basal cell diameter of 6-8 $\mu$m at the forearm and 
hand~\cite{corcuff1993morphometry}, at unexposed sites in the adult 
tissue~\cite{weinstein1984cell}, the average cell diameter in the proliferative 
compartment is about 10 $\mu$m whereas the average differentiated cell diameter 
is about 16 $\mu$m. For simplicity, we use ellipsoids to model geometric 
heterogeneity in cell morphology and size. Cell types are distinguished by the 
mean major-to-minor axis ratio and the mean nominal size.

\begin{table}[b]
\small
\centering
\caption{\label{tab:size} {\bf Parameters for cell size and the undulant rete ridge}}
\begin{tabular}{llc}\hline
 {\bf Cell type} & $w$ ($\mu m$) $\times$ $h$ ($\mu m$) &  {\bf Reference}\\\hline\hline
Stem cell & $10\times 15$ & \cite{weinstein1984cell,corcuff1993morphometry} \\
Transit-amplifying cell & $10\times 15$ & \cite{weinstein1984cell,corcuff1993morphometry} \\
Growth-arrested cell & $12\times 10$ & \cite{weinstein1984cell,corcuff1993morphometry} \\
Spinous cell & $12\times 8$ & \cite{weinstein1984cell,corcuff1993morphometry} \\
Granular cell & $22.5\times 4$ & \cite{weinstein1984cell,corcuff1993morphometry} \\
Cornified cell & $35\times 1$ & \cite{plewig1970regional,plewig1970regionald,kashibuchi2002three,marks1981studies} \\
\hline
\multicolumn{3}{l}{Rete ridge parameters (see Supplementary Materials):}\\
\multicolumn{2}{l}{$A=$63 $\mu$m, $B=$23 $\mu$m, $\sigma_s=$70 $\mu$m} & \cite{giangreco2009human} \\
\hline
\end{tabular}
\end{table}

The basement membrane of the epidermis is undulant with rete ridges extending 
downward between the dermal papillae [Fig.~\ref{fig:force}(c)]. In adult human 
epidermis, the average rete ridge height is about 40 $\mu$m in the adult tissue, 
and about six rete ridges along 1 mm cross-sectional tissue length were 
observed~\cite{giangreco2009human}, which changes with age. The basement 
membrane is modeled by periodically-repeating Gaussian functions (see the 
Supplementary Material). Parameters of cell sizes and epidermis thickness are 
listed in Table~\ref{tab:size}.

\begin{figure*}
\begin{center}
\includegraphics[scale=0.4]{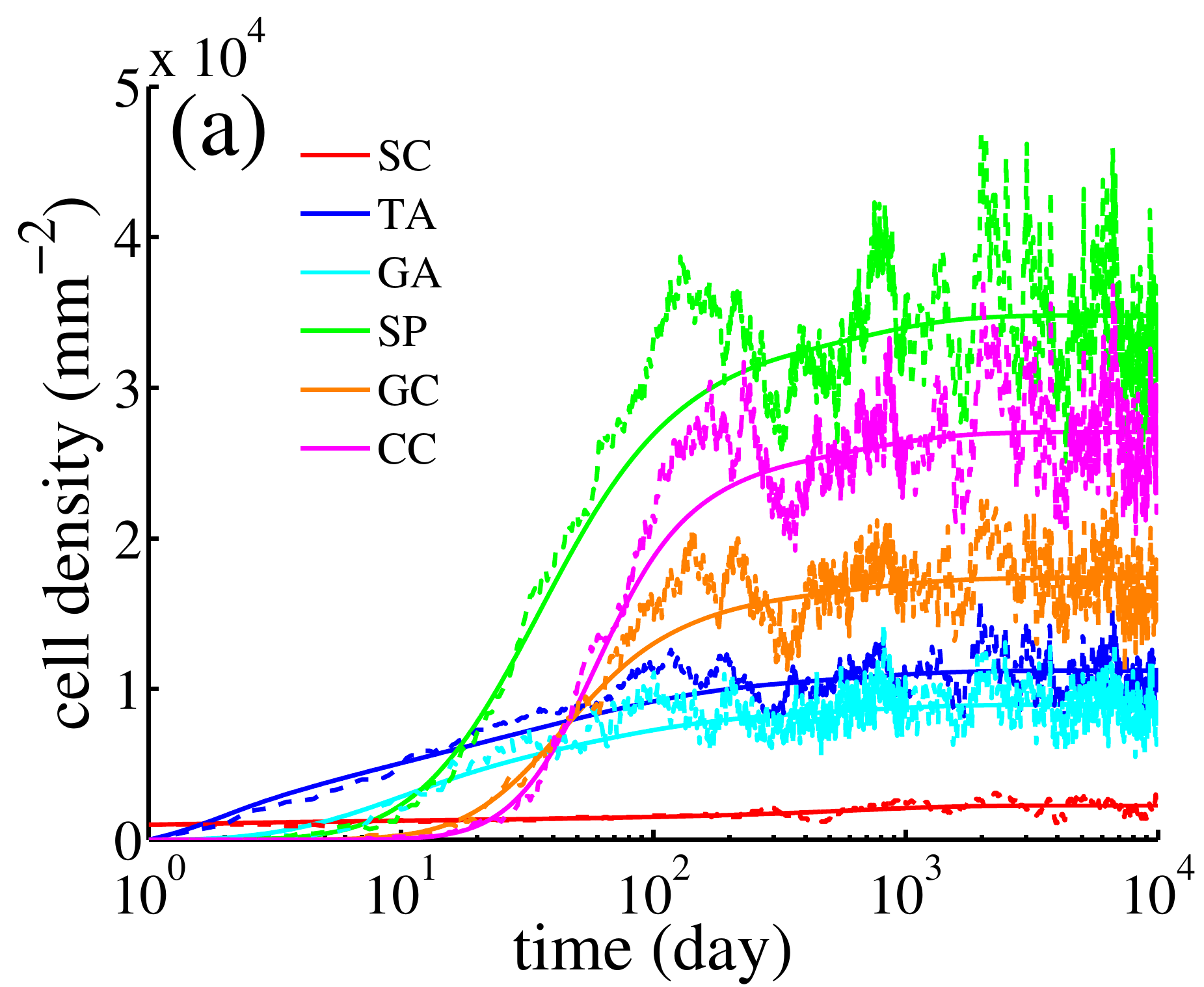}
\includegraphics[scale=0.4]{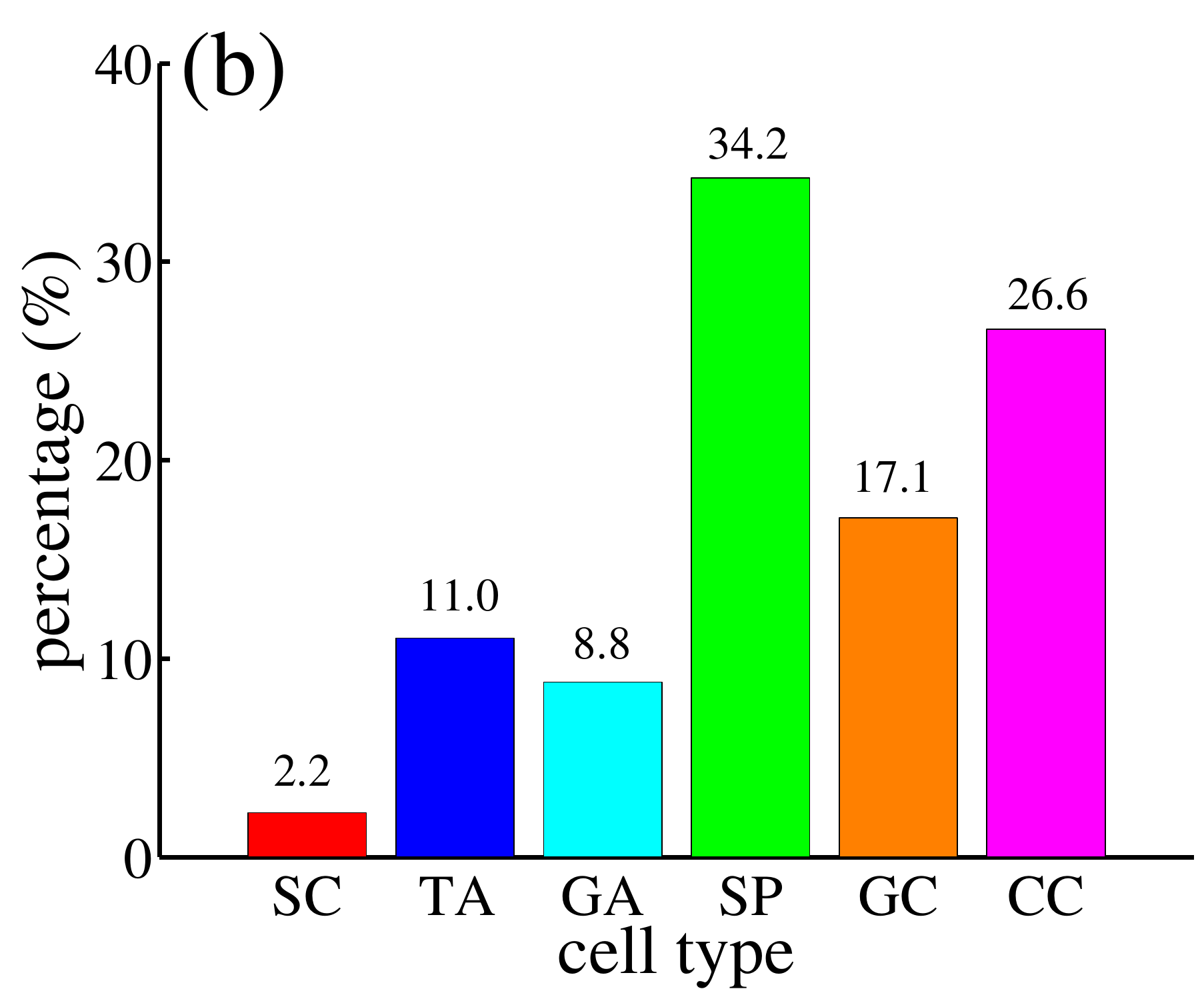}
\includegraphics[scale=0.21]{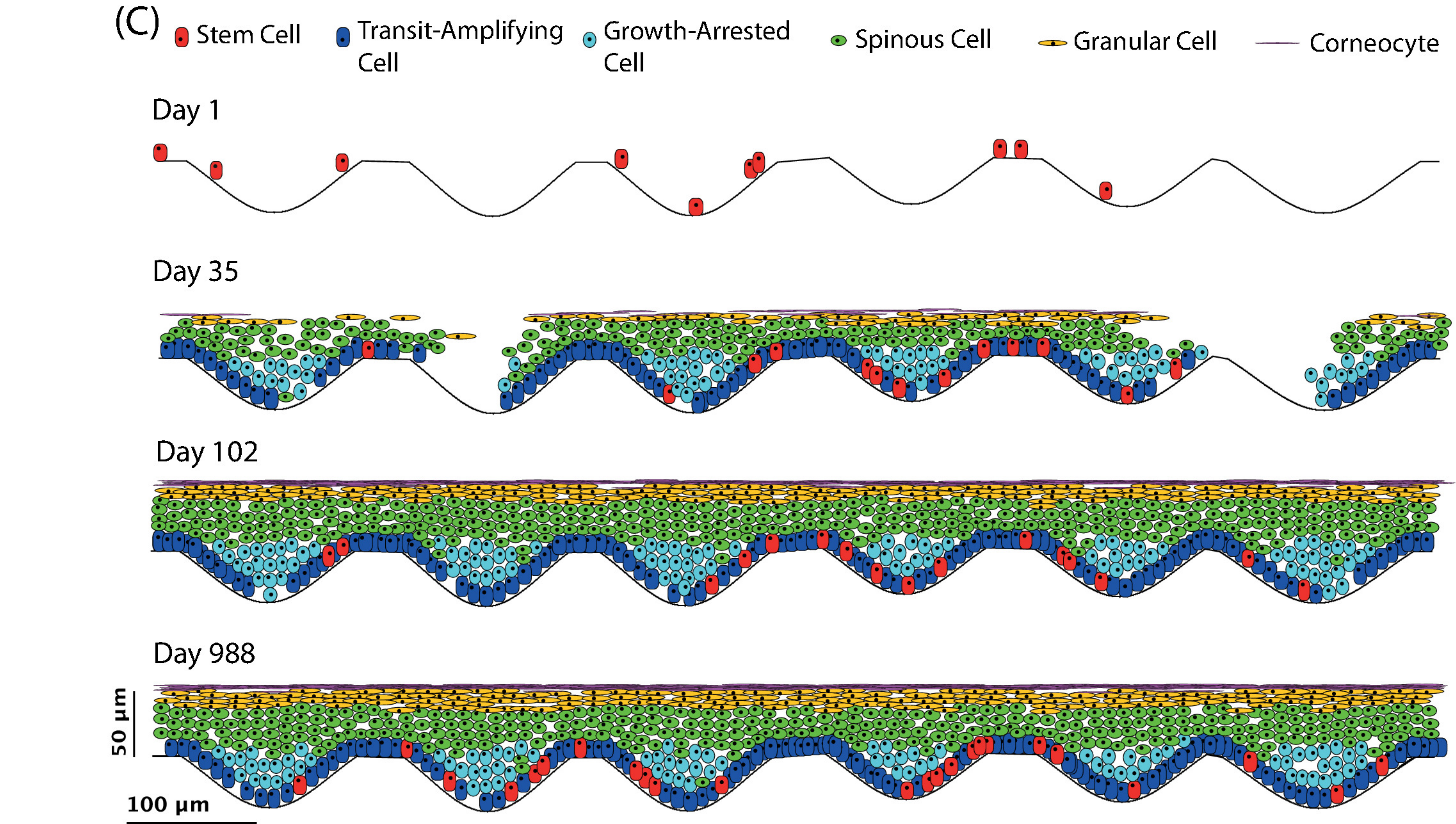}
\end{center}
\caption{\label{fig:odepopulation}{\bf Dynamic and homeostatic keratinocytes density distribution.} (a) Temporal evolutions of cell population of different types of keratinocytes by a deterministic simulation (smooth curves) and a stochastic simulation (fluctuated curves). Deterministic simulations started with an initial stem cell density of 1000 mm$^{-2}$, which corresponds to 10 cells under a skin area of 0.01 mm$^{-2}$ for the stochastic simulation. (b) The histogram of the steady-state cell density distribution. (c) Snapshots at day 1, 35, 102, and 988 of the visualization of the two-dimensional epidermis with a thickness of 10 $\mu$m. Simulation started with stem cells randomly located along the basement membrane (day 1). Parameter values used in simulations are listed in Table~\ref{tab:parameter}.}
\end{figure*}

\section{Results}

The model recapitulates two important measures of the epidermis homeostasis: 
cell counts in different layers and the epidermal turnover times in different 
compartments.

\subsection{Homeostatic cell density distribution}
The mean homeostatic cell densities can be analytically calculated from the 
ordinary differential equations in Table~\ref{tab:equations} as follows (see 
Table~\ref{tab:parameter} for definitions of parameters). 

\begin{equation}\label{eq:sc}
p_{sc} = p_{sc}^{\rm 
max}\left[1-\frac{1}{\gamma_1}\left(k_{1s}+\beta_1-\frac{k_{-1}(k_{1a}+2k_{1s})}
{k_{-1}+k_{2s}+\beta_2-\gamma_2-\frac{k_{-2}(k_{2a}+2k_{2s})}{k_{-2}+k_3+\beta_3
}}\right)\right] \ ,
\end{equation}

\begin{eqnarray}
p_{ta} & = & 
\frac{k_{1a}+2k_{1s}}{k_{-1}+k_{2s}+\beta_2-\gamma_2-\frac{k_{-2}(k_{2a}+2k_{2s}
)}{k_{-2}+k_3+\beta_3}}p_{sc} \label{eq:scta}\ ,\\
p_{ga} & =& \frac{k_{2a}+2k_{2s}}{k_{-2}+k_3+\beta_3}p_{ta}, \\
p_{sp}&=&\frac{k_3}{k_4+\beta_4}p_{ga} \ , \\
p_{gc}&=& \frac{k_4}{k_5+\beta_5}p_{sp}, \\\label{eq:cc}
p_{cc}&=&\frac{k_5}{\alpha}p_{gc} \ ,
\end{eqnarray}
and the total cell density is given as
\begin{equation}\label{eq:ptot}
p_{\rm tot}=p_{sc}+p_{ta}+p_{ga}+p_{sp}+p_{gc}+p_{cc} \ .
\end{equation}
The density of each cell category is proportional to the epidermis capacity of 
stem cells, $p_{sc}^{\rm max}$ and the ratio between densities of any pair of 
cell types is a constant. Therefore, observed cell density distribution can be 
used to identify kinetic parameters. The homeostatic TA-cell density is
\begin{equation}
p_{ta,h}=\frac{k_{1a,h}+2k_{1s,h}}{k_{-1}+k_{2s}+\beta_2-\gamma_2-\frac{k_{-2}
(k_{2a}+2k_{2s})}{k_{-2}+k_3+\beta_3}}p_{sc,h}
\end{equation}
Kinetic parameters must satisfy two necessary conditions to establish a 
physiologically proper steady state:
\begin{equation}
\gamma_{1,h}>\beta_1+k_{1s,h}-\frac{k_{-1}(k_{1a,h}+2k_{1s,h})}{k_{-1}+k_{2s}
+\beta_2-\gamma_2-\frac{k_{-2}(k_{2a}+2k_{2s})}{k_{-2}+k_3+\beta_3}}
\end{equation}
\begin{equation}
k_{-1}+k_{2s}+\beta_2>\gamma_2+\frac{k_{-2}(k_{2a}+2k_{2s})}{k_{-2}+k_3+\beta_3} 
\ .
\end{equation}
Given small backconversion rate ($k_{-1}$ and $k_{-2}$ assumed at $10^{-6}$ per 
day, 3-5 orders of magnitude smaller than the stem cell and TA cell proliferating 
rate constants, Table~\ref{tab:parameter}), the above conditions simplify to
\begin{equation}
\gamma_{1,h}>k_{1s,h}+\beta_1,  \ \  \text{and} \ \gamma_2<k_{2s}+\beta_2 \ ,
\end{equation}
The first condition ensures the establishment of a viable stem cell population, whereas the second condition prevents an unchecked growth of the epidermis because the proliferation of TA cells is not limited by a maximum capacity in the model. These conditions extend previous treatment that required a precise 
balance between self-proliferation and symmetric division in progenitor cells when apoptosis was neglected~\cite{clayton2007single,klein2007kinetics}. Apoptosis is a rare event compared to progenitor self-proliferation and symmetric division (e.g., $\beta_1\ll\gamma_{1,h}$, and $\ll k_{1s,h}$; see Table~\ref{tab:parameter}) and plays a secondary role in regulating the epidermal homeostasis.

A recent study of progenitors in mice interfollicular epidermis by Mascr{\'e} et al.~\cite{mascre2012distinct} suggested that in homeostasis stem cells proliferate by 4-6 division events per year, 10-20 times slower than proliferation and differentiation (about 1.2 events per week) of committed 
progenitors (TA cells in our model). This information is reflected such that $k_{1a,h}+2k_{1s,h}$ is an order of magnitude smaller than $\gamma_2+2k_{2s}$. However, considering that TA cells have a much larger population than stem cells ($p_{ta}/p_{sc}\approx 5$~\cite{heenen1997growth}), rates of stem-cell 
asymmetric and symmetric divisions into TA cells ($k_{1a,h}+2k_{1s,h}$) must be about 5-fold greater than the difference between TA-cell self-proliferation and symmetric division, i.e., $k_{2s}+\beta_2-\gamma_2$ based on Eq.~(\ref{eq:scta}), assuming backconversion rates ($k_{-1}$ and $k_{-2}$) are negligible. Therefore, the model predicts that mild perturbations on TA-cell dynamics including proliferation and apoptosis may have substantial effects on the homeostatic keratinocyte population.

The model recapitulates the dynamics and the homeostatic cell density distribution and the epidermal turnover time, in both deterministic and stochastic simulations [Fig.~\ref{fig:odepopulation}(a)]. Starting from an initial stem-cell population, the cell density rapidly increases to reach about 80\% of the homeostatic density in the transient phase of the first 100 days and significantly slows down when it approaches the homeostasis. The simulation reached the homeostasis in about 2000 days to a total density at 101684 cells/mm$^2$. The transient dynamics is more than an order of magnitude slower than that in the model by Grabe et 
al.~\cite{grabe2005multicellular,grabe2007simulating}, where the authors reported a near 2000-hour transient dynamics from an initial stem cell population to a homeostasis. This discrepancy is caused by a slow stem cell dynamics of 4-6 cell events per year reported recently~\cite{mascre2012distinct}, which also agrees with early-reported cell cycle of 100-200 hrs in a 20\% growth fraction in an animal model~\cite{morris1990epidermal}. Therefore, the stem cell dynamics (Eq.~\ref{eq:scrate}) becomes a rate-limiting step to the dynamic establishment of a homeostasis, notably when the system approaches the homeostasis.

Figure~\ref{fig:odepopulation}(b) shows a histogram of the simulated steady-state cell density distribution over different cell types. Stem cells, TA cells and GA cells in stratum basale consist of 22\% total cell population, in which proliferating cells accounts for 13.2\% total cell population, consistent 
with the experimental data~\cite{heenen1997growth,bauer2001strikingly,hoath2003organization}. Studies 
by Bergstresser et al.~\cite{bergstresser1978counting} also suggested that 30\% of nucleated keratinocytes were in the basal layer where stem cells occupied 10\% population~\cite{heenen1997growth}. Spinous and granular cells are the majority, consisting of 51.3\% total keratinocytes with the population of 
spinous cells nearly two times that of granular cells, in agreement with a previous prediction by Grabe et 
al.~\cite{grabe2005multicellular,grabe2007simulating}. Corneocytes consist of 26.6\% of the total cell population. Bauer et al.~\cite{bauer2001strikingly} measured a mean nucleated keratinocytes density of 75346 cells/mm$^2$ whereas the non-nucleated corneocytes was estimated to be about 18000 cells/mm$^2$, consisting of 19.3\% total cell population~\cite{hoath2003organization}.

Figure~\ref{fig:odepopulation}(c) shows snapshots of temporal evolution of two-dimensional epidermal stratification from an initial group of stem cells distributed along the basement membrane to the homeostasis. We compute the cell population dynamics and the cell migration within an area of 1 mm in length by 10 $\mu$m in width. The cell density is defined over a surface area number of keratinocytes per mm$^2$ without explicitly considering the epidermis height. The choice of 10 $\mu$m (about the mean cell size) is to visualize a two-dimensional single layer of keratinocytes. The simulated tissue histology shows that the thickness of the nucleated epidermis is about 60 $\mu$m, aligning with observations ranging from 38 $\mu$m to 77 $\mu$m with a mean of 60 $\mu$m in the adult tissue, with little variation across age 
groups~\cite{corcuff1993morphometry,giangreco2009human,bauer2001strikingly}. A movie of the epidermis renewal process of the normal tissue is available at URL: http://www.picb.ac.cn/stab/epidermal.html.

\subsection{The epidermal turnover time}
Another common measure of epidermis homeostasis is the epidermal turnover time $\tau$. At the tissue level,  $\tau$ is interpreted as a time required for replacing the entire epidermis with new keratinocytes. The epidermal turnover time varies significantly with age groups, tissue locations and cell densities. 
In earlier studies, $\tau$ has been reported approximately 6-7 weeks in the volar forearm with a nucleated cell density of 44000 
mm$^{-2}$~\cite{bergstresser1977epidermal,iizuka1994epidermal}. Based on a more recent count of nucleated cell density of 75346 mm$^{-2}$ on breast 
skin~\cite{bauer2001strikingly}, Hoath and Leahy~\cite{hoath2003organization} suggested that $\tau$ should be calculated as 59.3 days. Renewal of a specific layer of keratinocytes takes a shorter turnover time. The turnover times of the stratum basale and the differentiated compartment were reported being about 22 
and 12 days~\cite{iizuka1994epidermal}, whereas the stratum corneum has a 
turnover time that varies from 14 days~\cite{
iizuka1994epidermal,weinstein1965autoradiographic} to about 20 days in young 
adult~\cite{grove1983age}. Following the convention~\cite{iizuka1994epidermal}, 
we calculate $\tau$ as the ratio of total cell density to the rate of cell loss 
including desquamation and apoptosis, or alternatively, the ratio of total cell 
density to the rate of cell birth by stem cell and TA cell divisions because 
cell birth and death rates are balanced at homeostasis:
\begin{equation}
\tau = \frac{p_{\rm tot}}{r_{\rm growth}} =\frac{p_{\rm tot}}{r_{\rm loss}} \ 
,\label{eq:turno}
\end{equation}
where the rates of cell growth and cell loss are
\begin{eqnarray}
r_{\rm growth}&= &[\gamma_1(1-p_{sc}/p_{sc}^{\rm max})+k_{1a}+k_{1s}]p_{sc} 
\notag\\
 &&+(\gamma_2+k_{2a}+k_{2s})p_{ta}\notag \\
r_{\rm loss}&=& 
\beta_1p_{sc}+\beta_2p_{ta}+\beta_3p_{ga}+\beta_4p_{sp}+\beta_5p_{gc}+\alpha 
p_{cc} \notag \ .
\end{eqnarray}
Analytical solution to $\tau$ can be obtained by substituting Eqs.~(\ref{eq:sc})-(\ref{eq:cc}) into the above equation. Apoptosis events represent rare ramifications from the central transition pathway in the normal 
epidermis renewal. Rate constant of an apoptotic process $\beta_i$ (10$^{-4}$-10$^{-5}$/day) is much smaller than the differentiation and desquamation rate constants $k_i$ and $\alpha$ (about $10^{-1}$/day, see 
Table~\ref{tab:parameter}). We can approximate $r_{\rm loss}$ by the rate of desquamation. By also neglecting the backconversions, we have
\begin{equation}\label{eq:to}
\tau=\tau_{\rm prolif}+\tau_{\rm diff}+\tau_{\rm corn} \ ,
\end{equation}
where
\begin{eqnarray}
\tau_{\rm prolif}&=&\left(\frac{k_{2s}-\gamma_2}{k_{1a}+2k_{1s}} 
+1\right)\frac{1}{k_{2a}+2k_{2s}} \label{eq:tp}\\
\tau_{\rm diff}&=&\frac{1}{k_3}+\frac{1}{k_4}+\frac{1}{k_5} \notag\\
\tau_{\rm corn}&=&\frac{1}{\alpha} \notag \ .
\end{eqnarray}
$\tau$ is the sum of contributions by sub-compartment turnover times as keratinocytes migrate from the proliferative compartment ($\tau_{\rm prolif}$) through the differentiated compartment ($\tau_{\rm diff}$) and then through the stratum corneum ($\tau_{\rm corn}$). We note that $\tau$ is independent of the 
stem cell self-proliferation $\gamma_1$ and the stem cell capacity $p_{sc}^{\rm max}$. These two parameters determine the steady-state stem cell density $p_{sc}$. Both cell density and cell growth rate are proportional to $p_{sc}$, which masks the effect of stem cell self-proliferation dynamics on the turnover 
times. The turnover time of a single cell category is the inverse of the rate constant for the transit to the next cell category. Therefore, $\tau$ can be dominated by a rate-limiting step in the cell growth and differentiation cascade. From the calculation by Eq.~(\ref{eq:turno}) using parameters in 
Table~\ref{tab:parameter}, $\tau$ is about 52.5 days, partitioned into 7, 31.5 and 14 days in the 
proliferative compartment, differentiated compartment and the stratum cornuem, respectively, within the wide range of reported adult-tissue measurements between 39 and 75 
days~\cite{weinstein1984cell,bergstresser1977epidermal,
weinstein1965autoradiographic}.

\begin{table*}[t]
\small
\centering
\caption{ \label{tab:parameter}{\bf Model parameters}}
\begin{tabular}{llll}
\hline
 \multicolumn{4}{l}{\bf Cell population kinetics
model}\\\hline\hline
{Parameter description} &{Notation (unit)} & {Value}  & {Source}\\\hline
SC growth capacity & $p_{sc}^{\rm max}$ (mm$^{-2}$) & 4.50e3 & Assumed \\
Nominal SC self-proliferation rate constant & $\gamma_{1,h}$ (d$^{-1}$)&3.30e-3 
& Estimated \\
Nominal symmetric SC division rate constant  &$k_{1s,h}$ (d$^{-1}$) &1.64e-3 
&\cite{clayton2007single,mascre2012distinct} \\
Nominal asymmetric SC division rate constant &$k_{1a,h}$ (d$^{-1}$) &1.31e-2 
&\cite{clayton2007single,mascre2012distinct} \\
TA cell self-proliferation rate constant &$\gamma_{2}$ (d$^{-1}$) 
&1.40e-2&\cite{heenen1997growth,bauer2001strikingly,hoath2003organization,
bergstresser1978counting} \\
TA cell symmetric division rate constant & $k_{2s}$ (d$^{-1}$) & 1.73e-2 
&\cite{heenen1997growth,bauer2001strikingly,hoath2003organization,
bergstresser1978counting} \\
TA cell asymmetric division rate constant & $k_{2a}$ (d$^{-1}$) 
&1.38e-1&\cite{heenen1997growth,bauer2001strikingly,hoath2003organization,
bergstresser1978counting} \\
GA-to-SP cell differentiation rate constant & $k_{3}$ (d$^{-1}$) &2.16e-1 & 
\cite{bauer2001strikingly,hoath2003organization}\\
SP-to-GC cell transit rate constant$^\dagger$ & $k_{4}$ (d$^{-1}$) &5.56e-2 & 
\cite{bauer2001strikingly,hoath2003organization}\\
GC-to-CC cell transit rate constant$^\dagger$ & $k_{5}$ (d$^{-1}$) &1.11e-1 & 
\cite{bauer2001strikingly,hoath2003organization}\\
CC cell desquamation rate constant & $\alpha$ (d$^{-1}$) &7.14e-2 
&\cite{bauer2001strikingly,hoath2003organization,weinstein1965autoradiographic}
\\
Backconversion rate constant (TA to SC) & $k_{-1}$ (d$^{-1}$) &1.00e-6& Assumed 
\\
Backconversion rate constant (GA to TA) & $k_{-2}$ (d$^{-1}$) &1.00e-6& Assumed 
\\
Maximum fold increase of SC proliferation rate & $\omega$ & 100 & 
\cite{heenen1998ki,heenen1997growth,morris1983epidermal} \\
Steepness SC proliferation rate regulation by TA cell population & $n$ & 3 & 
Assumed \\
Normal epidermal apoptosis index     & AI$_{h}$ & 0.12\% 
&\cite{laporte2000apoptosis,bauer2001strikingly,hoath2003organization} \\
Psoriatic epidermal apoptosis index   &  AI$_d$ & 0.035\% & 
\cite{laporte2000apoptosis,bauer2001strikingly,hoath2003organization} \\
Fold change of psoriatic SC proliferation  & $\rho_{sc}$& 4 
&~\cite{weinstein1985cell}\\
Fold change of psoriatic TA proliferation & $\rho_{ta}$ &4 
&~\cite{weinstein1985cell} \\
Fold change of psoriatic cell transit rate $^\dagger$ & $\rho_{tr}$ & 5 
&~\cite{weatherhead2011keratinocyte,weinstein1973cytokinetics}\\
Fold change of psoriatic corneocyte desquamation & $\rho_{de}$& 4 
&~\cite{weinstein1965autoradiographic}\\
Fold change of psoriatic SC growth capacity & $\lambda$ & 3.5 
&~\cite{heenen1987psoriasis,simonart2010epidermal}\\
Maximum immune killing rate & $K_p$ ($\text{mm}^{-2}$d$^{-1}$)& 6 & Assumed\\
Immune half-activation psoriatic SC density & $K_a$ (mm$^{-2}$) & 380 & 
Assumed\\\hline
\multicolumn{4}{l}{\bf Cell migration model} \\\hline\hline
Parameter description & {Notation (unit)} & Value &   {Source}\\ \hline
Viscosity coefficient & $\mu$ (nN$\cdot$s$\cdot$$\mu$m$^{-1}$) & 250  & 
~\cite{palsson2008} \\
Elasticity constant & $k$ (nN$\cdot$$\mu$m$^{-2}$) & 0.04  &  Estimated \\
Adhesion factor of proliferating cells, SC and TA & $\sigma_p$ (nN)& 50 
&~\cite{li2013skin,palsson2008}\\
Adhesion factor of GA and SP and GC & $\sigma_d$ (nN)& 5 
&~\cite{palsson2008,li2013skin,skerrow1989changes} \\
Adhesion factor of CC&$\sigma_c$ (nN)& 0.5 
&~\cite{palsson2008,li2013skin,skerrow1989changes} \\
Mean cell radius & $r$ ($\mu$m) & 5 & Estimated \\
Time constant of rete ridge remodeling & $\tau$ (d) & 1000 & 
~\cite{Dawe1998,trehan2002,Hartman2002}\\
Maximum rete ridges height of psoriatic epidermis&$Y_{\rm max}$ ($\mu$m)&$126$& 
Assumed \\
Minimal rete ridges height of normal epidermis&$Y_{\rm min}$ 
($\mu$m)&40&~\cite{giangreco2009human}\\\hline
\multicolumn{4}{l}{$^\dagger$In the psoriasis model, due to lack of the granular 
layer $k_4$ is the rate constant for SP to CC transition,} \\
\multicolumn{4}{l}{$k_5$ is unused, and $\rho_{tr}$ is the fold change in $k_3$ 
and $k_4$.} 
\end{tabular}
\end{table*}

\subsection{Pathogenesis of psoriasis}

To investigate an important epidermis disorder, we extend the above model to 
study the onset and recurrence of psoriasis and its management. Psoriasis, an 
immune system-mediated chronic skin condition, is characterized by an 
overproduction of keratinocytes accompanied by 
inflammation~\cite{lowes2007pathogenesis}, resembling features found in many 
autoimmune diseases. Here, we focus on studying the most prominent type of the 
disorder, psoriasis vulgaris, a plaque-formed scaly silvery patch, affecting the 
majority of psoriasis patients. 

Current studies are inconclusive about whether psoriasis (1) arises from 
genetically hyperproliferative progenitor keratinocytes, or (2) is alternatively 
induced by a faulty immune system, particularly dendritic cells and T 
lymphocytes over-response to unresolved self antigens, which in turn produces 
excessive cytokines (such as TNF$\alpha$) to promote keratinocyte proliferation, 
or (3) more likely an intricate interplay between keratinocytes and the immune 
system~\cite{lowes2007pathogenesis,griffiths2007pathogenesis}.

\begin{figure}
\centering
\includegraphics[scale=0.3]{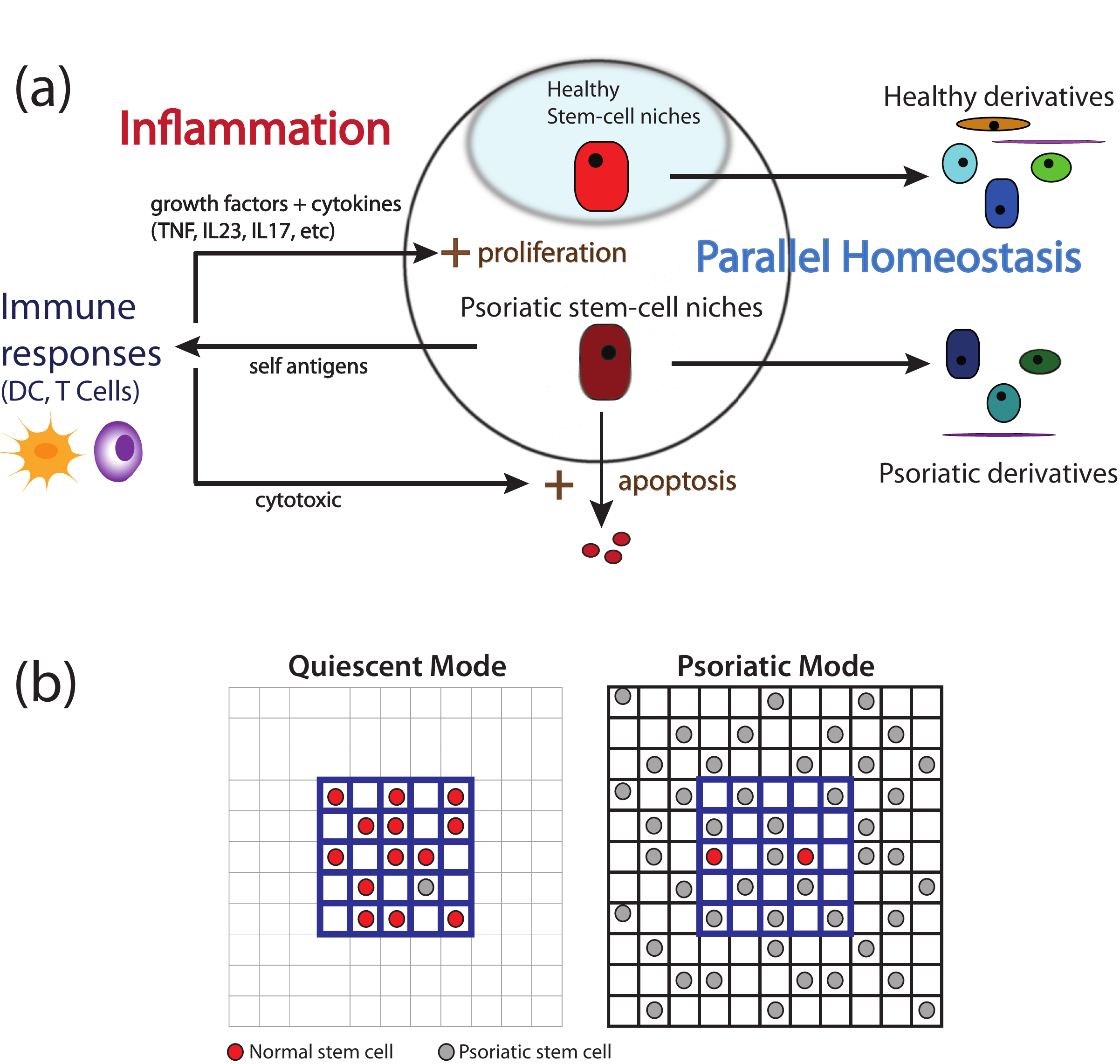}
\caption{\small {\bf Illustration of stem-cell niche environments in the psoriatic tissue}. \label{fig:niche} At the quiescent mode (left panel), healthy stem cells dominate in population, residing in the normal niche repertoire. At the psoriatic mode, psoriatic stem cells dominate in population and the niche repertoire expands to accommodate a larger population of psoriatic stem cells.}
\end{figure}

\begin{table*}[t]
\small
\centering
\caption{ \label{tab:distro}{\bf Cell density distribution ($\rm{mm^{-2}}$) and turnover times (day)$^*$}}
\begin{tabular}{p{3.5 cm}p{3 cm}p{3 cm}p{3 cm}}\hline
&Healthy tissue&Psoriasis$^\dagger$ &  Non-symptom$^\dagger$  \\ \hline\hline
{SC}&2268&362+6459&2232+124\\
{TA} &11219&77+32098&10715+618\\
{GA} &8964&61+20536&8562+395\\
{SP} &34799&238+79788&33236+1536\\
{GC} &17379&119+0&16598+0\\
{CC} &27055&185+77633&25840+1495\\
{Total} &101684&1042+216514&97183+4168\\
{$\tau$=$\tau_{\rm prolif}$+$\tau_{\rm diff}$+$\tau_{\rm corn}$}&52.5=7.0+31.5+14.0&9.8=1.8+4.5+3.5& 44.5=6.0+26.5+12.0\\\hline
\multicolumn{4}{l}{$^\dagger$Normal+Psoriatic ($p_{x}+\tilde{p}_{x}$).}\\
\multicolumn{4}{l}{$^*$Results are generated using parameters in Table~\ref{tab:parameter}.}
\end{tabular}
\end{table*}

To examine the pathogenesis of psoriasis, we emphasize an interplay between 
keratinocytes overproduction and responses by the immune system. Our fundamental 
hypothesis is that a psoriatic epidermis assumes two groups of keratinocytes: 
normal and psoriatic, maintaining a parallel homeostasis. The psoriatic 
keratinocytes are derived from a hyperproliferative stem cell population 
co-residing with the normal stem cells at the basement membrane, which compete 
for limited stem cell niches ((see Fig.~\ref{fig:niche} for illustration). This 
hypothesis can be justified by the existence of intrinsically hyperproliferative 
stem cells or stem cells that are more responsive to growth stimulants 
originated from an activated immune system. We modify the above model to 
describe the competition for niches between the normal and psoriatic stem cells. 
Dynamics of psoriatic TA and nonproliferative cells are governed by rate 
equations as in Table~\ref{tab:equations} for the normal keratinocytes, however, 
with different parameter 
values to generate known phenotypes in psoriatic plaques, with an exception of 
granular cells that are missing in psoriasis. The dynamics of psoriatic stem 
cells in our model is similar to the model of spruce budworm outbreak by Ludwig 
et al.~\cite{ludwig1978qualitative}, which models a single-species population 
growth under predation.
\begin{equation}\label{eq:ps1}
\frac{dp_{sc}}{dt}=\left[\gamma_1\left(1-\frac{p_{sc}+\tilde{p}_{sc}/\lambda}{p_
{sc}^{\rm max}}\right)-k_{1s}-\beta_1\right]p_{sc}+k_{-1}p_{ta}
\end{equation}
\begin{eqnarray}\label{eq:ps2}
\frac{d\tilde{p}_{sc}}{dt}& = 
&\left[\rho_{sc}\gamma_{1,h}\left(1-\frac{p_{sc}+\tilde{p}_{sc}}{\lambda 
p_{sc}^{\rm 
max}}\right)-\rho_{sc}k_{1s,h}-\tilde{\beta}_{1}\right]\tilde{p}_{sc} \notag \\
&&-f(\tilde{p}_{sc})+\tilde{k}_{-1}\tilde{p}_{ta}
\end{eqnarray}
\begin{equation}\label{eq:ps3}
 f(\tilde{p}_{sc})=\frac{K_p\tilde{p}_{sc}^2}{K_a^2+\tilde{p}_{sc}^2} \ .
\end{equation}
The psoriatic tissue activates the immune system to combat disease cells. We 
assume that repertoires for normal $p_{sc}$ and psoriatic $\tilde{p}_{sc}$ stem 
cells are both limited by available niche environment and psoriatic stem cells 
can acquire a larger growth capacity. Parameter $\lambda$ ($>1$) accounts for 
the fold increase in the growth capacity accessible to stem cells. The density 
of normal stem cells remains limited by $p_{sc}^{\rm max}$, which is invaded by 
a fraction ($1/\lambda$) of psoriatic stem cells. Equation~(\ref{eq:ps3}) models 
the immune activities triggered by psoriatic stem cells. An activated immune 
system induces apoptosis of psoriatic stem cells. The activity of the immune 
system (the killing rate, $f(\tilde{p}_{sc})$) is directly regulated by the 
psoriatic stem cell density, under the assumption that the immune system is 
activated in a faster time scale than the tissue growth. The immune response is 
significantly activated when $\tilde{p}_{sc}$ exceeds a threshold parameterized 
by $K_a$ and is saturated at the maximum rate $K_p$ at $\tilde{p}_{sc}\gg K_a$ 
when the psoriatic stem-cell population overwhelms that of cytotoxic T cells. 
This approach hypothesizes that the immune system combats disease stem cells, 
but does not exclude the commonly-believed role by the immune system of 
inducing keratinocyte overproduction, even though the model does not explicitly 
couple the immune system to stem-cell proliferation.

Proliferation rate constants $\gamma_1$, $k_{1s}$ and $k_{1a}$ for the normal 
stem cells are regulated by the total TA cells, $p_{ta}+\tilde{p}_{ta}$, similar 
to Eq.~(\ref{eq:scrate}):
\begin{equation}\label{eq:scrateps}
\frac{\gamma_1}{\gamma_{1,h}}=\frac{k_{1a}}{k_{1a,h}}=\frac{k_{1s}}{k_{1s,h}}
=\frac{\omega}{1+(\omega-1)[(p_{ta}+\tilde{p}_{ta})/p_{ta,h}]^n} \ .
\end{equation}
In comparison, we assume that psoriatic stem cells are not subject to regulation 
by the TA cell population and proliferate with rates $\rho_{sc}$-fold higher 
than the homeostatic rate constants ($\gamma_{1,h}$, $k_{1a,h}$ and $k_{1s,h}$) 
of normal stem cells. We assume that the immune response substantially switches 
on when the psoriatic stem cell population reaches 10\% of stem cell population 
in the normal tissue. This assumption is used to parameterize the steepest 
change in the removal rate $f(\tilde{p}_{sc})$ in response to $\tilde{p}_{sc}$, 
which sets the half-activation density at $K_a=\sqrt{3}p_{sc,h}/10$. The model 
does not consider immune responses against cells derived from psoriatic stem 
cells by observing that reduction in stem cell population results in a 
subsequent reduction in the derived keratinocyte population.

Despite its mechanistic uncertainties, psoriasis plaques have well-defined 
tissue-level phenotypes, making it a good candidate for study by a predictive 
model. Depending on its severity a plaque exhibits 2-5 times increase in the 
total cell 
density~\cite{weinstein1985cell,weatherhead2011keratinocyte,van1963kinetics} 
with a relatively higher growth in the proliferative compartment compared to the 
nonproliferative compartment~\cite{heenen1987psoriasis,simonart2010epidermal}. A 
disordered tissue usually loses the granular layer due to abnormal 
differentiation and contains a subset of nucleated corneocytes. A psoriatic 
plaque also has a turnover time several fold 
faster~\cite{weinstein1965autoradiographic,lowes2014immunology}. More 
specifically, studies found in the cell kinetics of psoriasis (i) significant 
increase of cell cycle marker Ki-67 in psoriatic tissue without much change in 
cell cycle time~\cite{castelijns2000cell,doger2007nature}, suggesting a 
substantial increase in growth fraction; (ii) transit time of keratinocytes 
through differentiated compartment is shortened to 48 hrs from 240-330 hrs, 5-7 
times faster than in the normal tissue~\cite{weinstein1973cytokinetics}; (iii) 
transit time through the corneum is also shortened from 14 days to 2 
days~\cite{weinstein1965autoradiographic}. These factors together result in a 
decrease in the epidermis turnover time [Eq.~(\ref{eq:to})]. In addition, the 
cell apoptotic index decreases nearly four fold from 0.12\% to 0.035\%~\cite{laporte2000apoptosis}, making a further contribution to keratinocytes overproduction. Table~\ref{tab:parameter} lists parameter values 
based on the above observations, where coefficients $\rho_{ta}$, $\rho_{tr}$ and 
$\rho_{de}$ are fold changes over rate constants in normal kinetics of TA-cell 
proliferating ($\gamma_2$, $k_{2a}$ and $k_{2s}$), transit in the 
nonproliferative compartment ($k_3$ and $k_4$) and desquamation ($\alpha$), 
respectively. Like variations in normal tissues, severity and phenotype of 
psoriasis vary widely across individuals and disease subtypes and therefore for any specific condition or 
study the model should be parameterized accordingly.

\begin{figure*}
\centering
\includegraphics[scale=0.5]{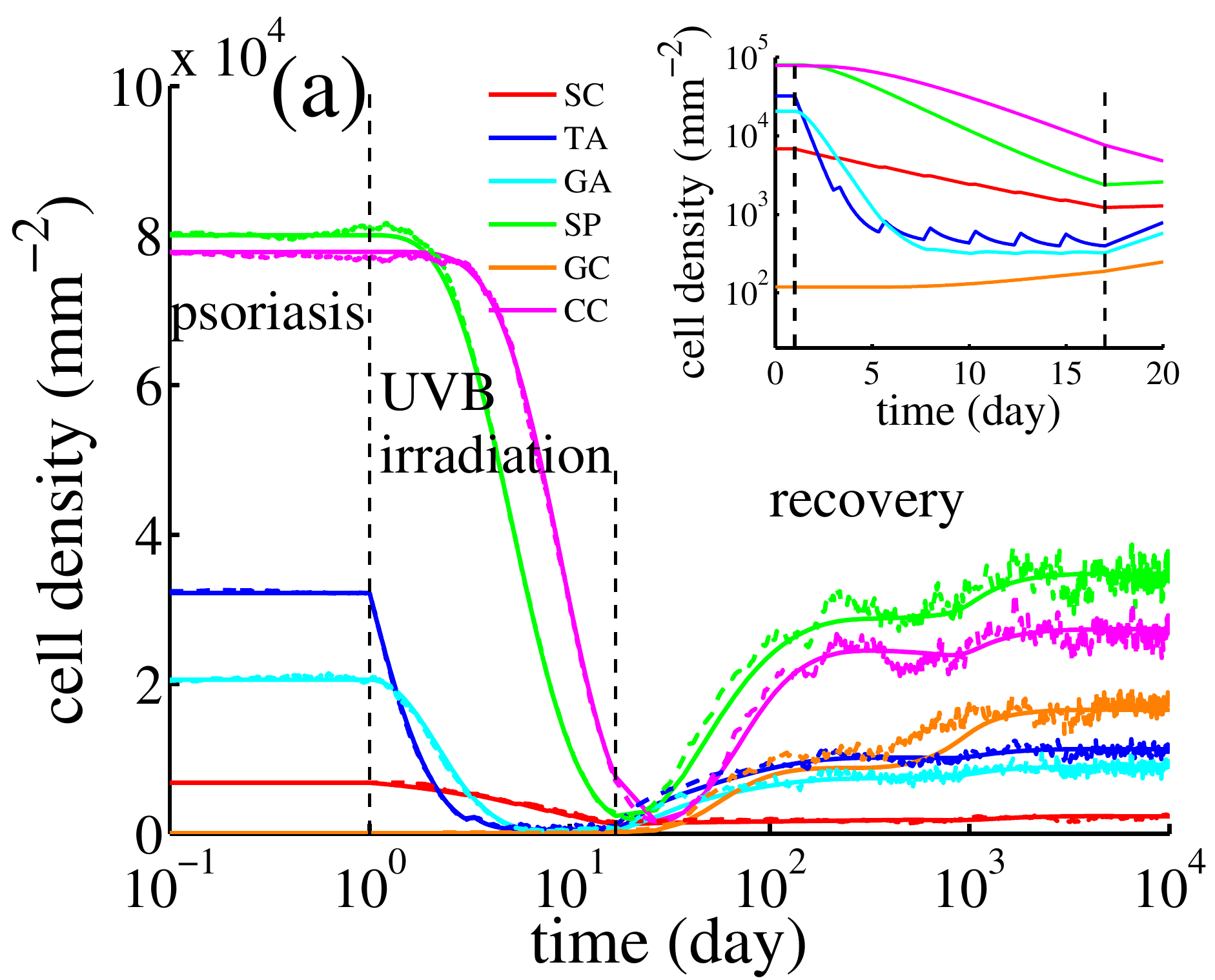}

\includegraphics[scale=0.33]{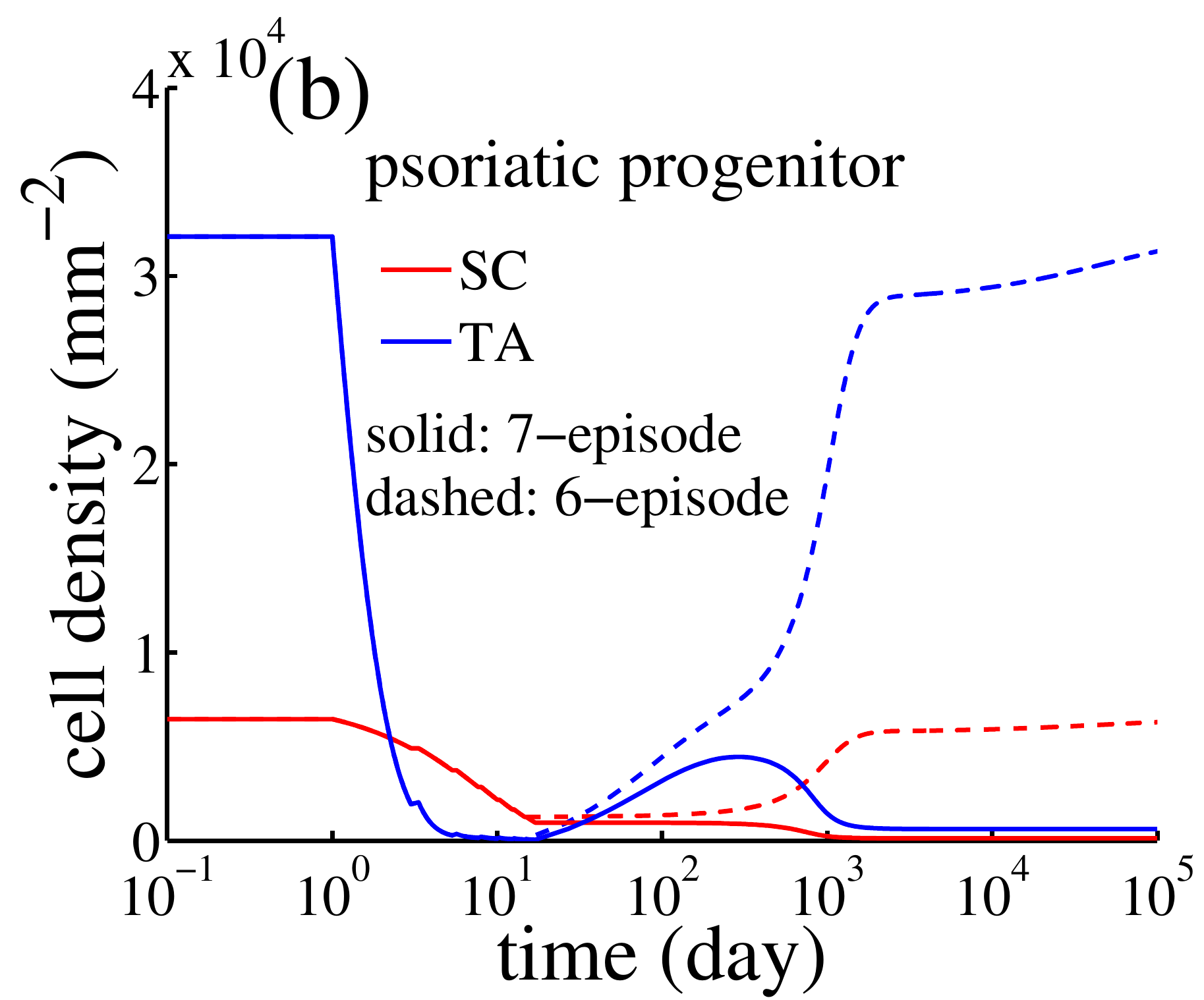}
\includegraphics[scale=0.33]{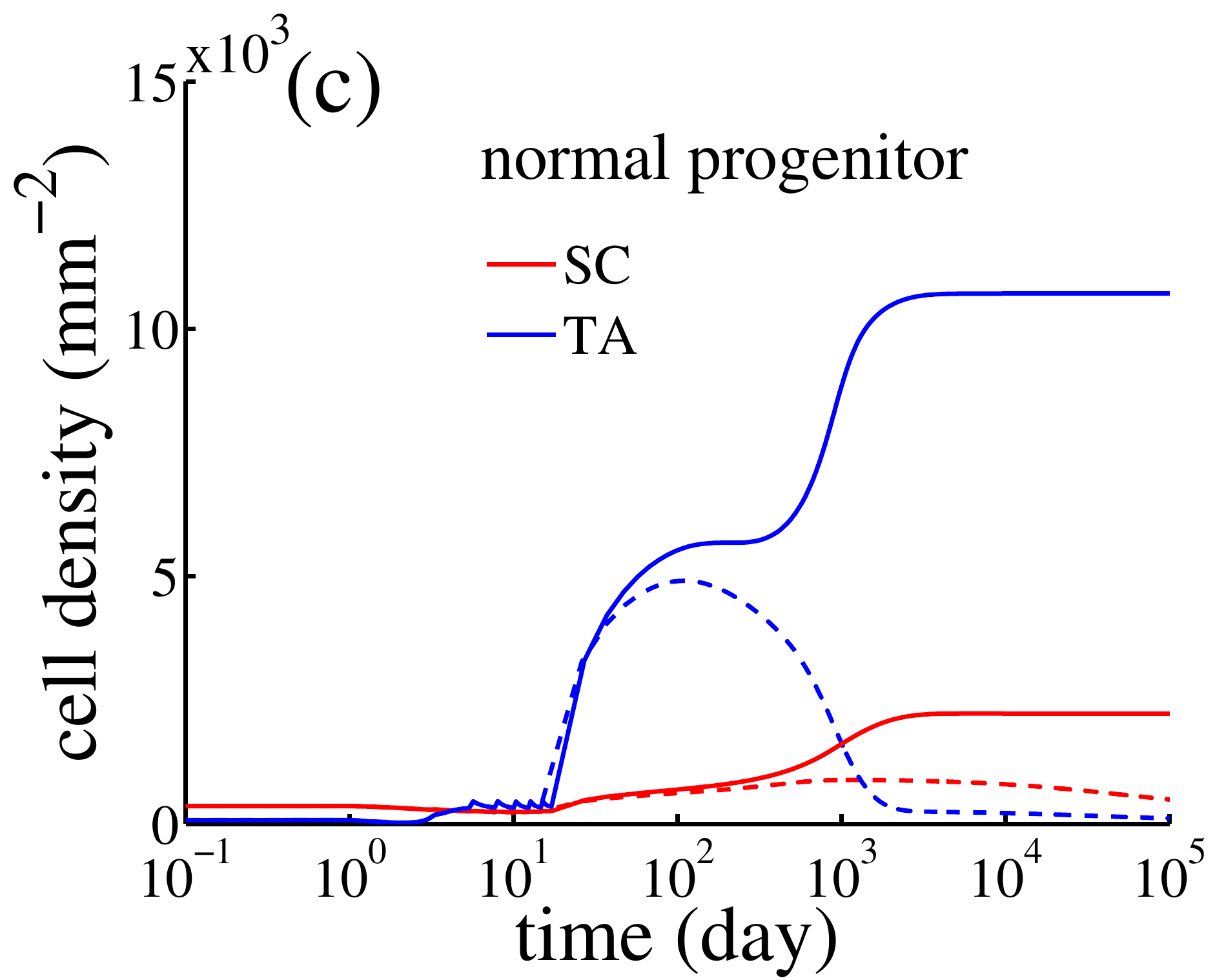}

\includegraphics[scale=0.33]{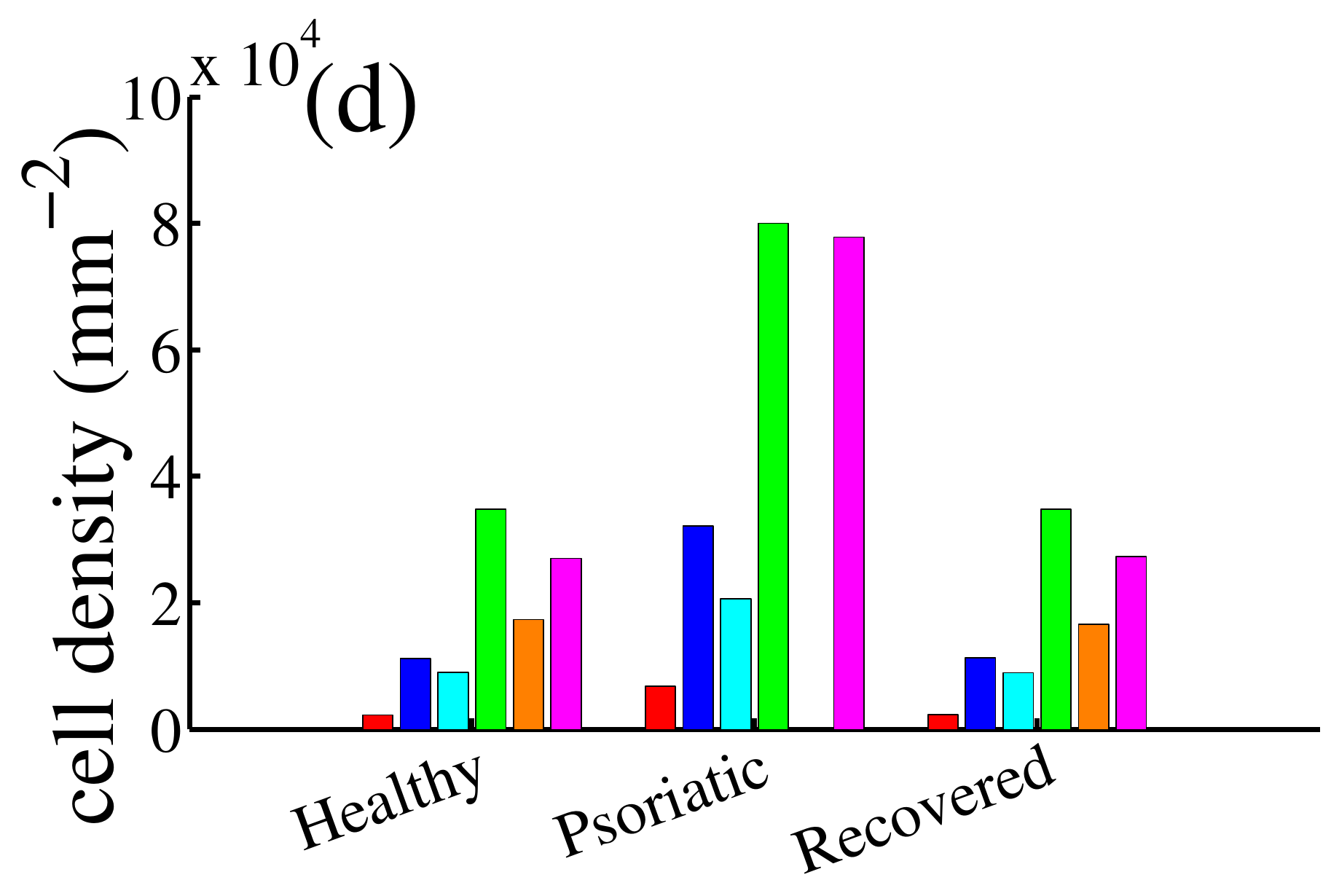}
\includegraphics[scale=0.33]{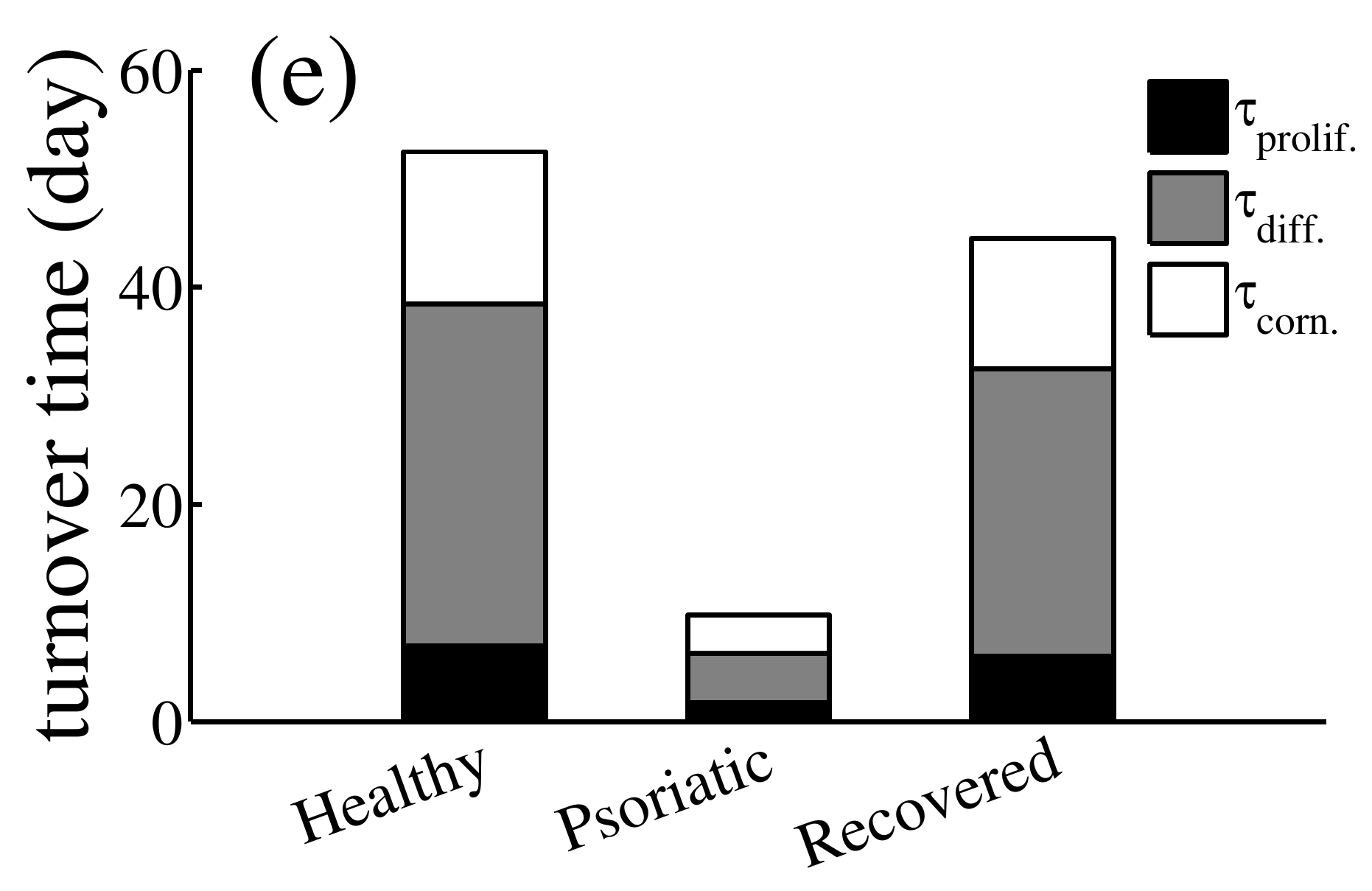}
\caption{{\bf Psoriasis and its management by phototherapy.} (a) Three dynamical 
phases of cell densities from psoriasis, the UVB treatment and recovery. 
Trajectories are combined populations of psoriatic and normal cells. Both 
stochastic and deterministic trajectories are shown. The stochastic simulation 
was conducted in an area of 0.1 mm$^{-2}$. Inset: zoomed details of the UVB 
irradiation phase (from day 1 to day 17). Psoriatic (b) and normal (c) stem cell 
and TA cell densities before and after 6- or 7-episode treatment. The treatment 
started at time day 1. Each simulated UVB irradiation episode induces 60000-fold 
increases in apoptosis rate constants for stem cells and TA cells and lasts for 
48 hours. The time interval between two consecutive episodes is 56 hours (48-hr 
irradiation + 8-hr resting). (d) and (e) Homeostatic cell density distributions 
and the epidermal turnover times for healthy, psoriatic and recovered tissues. 
Notice that the granular layer is missing in the psoriatic tissue.}
\label{fig:recover}
\end{figure*}

As the main result, the model predicts two interconverting homeostatic modes of 
the psoriatic tissue (see the Appendix for analytical details): (1) a {\em 
disease} state, which generates psoriasis phenotypes of keratinocytes 
overproduction and a shortened epidermal turnover time when psoriatic stem cells 
outcompete normal stem cells for available niches and overwhelm the immune 
system; and (2) a {\em quiescent} state, which predicts a coexistence of a small 
number of psoriatic cells with a dominant population of normal keratinocytes 
when the immune system keeps the psoriatic stem cell population low. The 
quiescent mode with remanent psoriatic stem cells is a symptomless state for a 
psoriasis susceptible tissue and may relapse into the disease state given 
favorable conditions. On the other hand, the disease state may be reverted back 
to the quiescent state by a properly designed treatment.

The model is parameterized (Table~\ref{tab:parameter}) to produce a specific 
psoriasis phenotype, in which the total cell density reaches 217000 mm$^{-2}$ at 
homeostasis, more than two times the normal epidermis (101600 mm$^{-2}$), with 
the pathological keratinocytes consisting of 99.5\% of the total population. The 
relative growth of the proliferative compartment over the nonproliferative 
compartment is about 3 vs. 2 times the normal densities 
(Table~\ref{tab:distro}). The epidermal turnover time is shortened more than 5 
times from 52.5 days to 9.8 days.

The psoriatic epidermis may retreat to the ``quiescent" state and achieve a 
remission provided that the psoriatic stem cell density can be managed below a 
threshold value (see the Appendix). Narrow-band 311 nm controlled UVB 
irradiation is known as an effective treatment for managing 
psoriasis~\cite{menter2007current}. For example, a recent study by Weatherhead 
et al.~\cite{weatherhead2011keratinocyte} applied sequential episodes of 0.75-3 
MEDs (minimal erythemal dose) UVB irradiation to achieve plaques remission by 
inducing strong apoptosis of proliferative cells. Figure~\ref{fig:recover}(a) 
shows a model simulation of a psoriasis remission after a simulated sequence of 
7-episode UVB irradiations. Each UVB irradiation episode is simulated by 
increasing apoptosis rate constants 60000 fold indiscriminately for normal and 
psoriatic stem cells and TA cells for 48 hours followed by a 8-hour resting 
interval before starting the next episode. The entire treatment lasts 16 days. 
Model simulations unveil an intriguing interplay between dynamics of psoriatic 
and normal cells. Upon the initiation of irradiation, the total cell population 
first declines due to the UVB-induced apoptosis in stem cells and TA cells. 
Each episode of UVB irradiation induced apoptosis in about 22\% stem cells in 
the pre-episode population [see Fig.~\ref{fig:recover}(a) inset]. The 
population of TA cells declined more substantially due to combined effects of 
increased apoptosis and reduced stem-cell symmetric and asymmetric divisions. A 
mild rebound of cell densities happens in each resting interval because of a 
continuing hyper-proliferation of psoriatic stem cells and TA cells. At the end 
of the treatment, the total keratinocyte density dramatically drops more than 
95\% from 217000 to 12100 mm$^{-2}$, with normal and psoriatic stem cells 
respectively reduced to 260 and 960 mm$^{-2}$. Post-treatment stem-cell 
population continues to decline to the ``quiescent" steady state due to a 
relatively stronger immune response [Fig.\ref{fig:recover}(b) and Fig.~S3(c) 
and related text in the Supplementary Materials], which later brings down the 
psoriatic cell density to a minimum (less than 0.5\% of the total cell density, 
Table~\ref{tab:distro}). The granular cells become visible during the recovery. 
The last phase indicates a recovery of keratinocytes derived from the normal 
stem cells that reclaim their niche repertoire by a slower kinetics 
[Fig.~\ref{fig:recover}(c)]. The downstream differentiated cells follow similar 
dynamics of remission. Simulated dynamics is similar if the model considers 
UVB-induced apoptosis in all nucleated cells [results not shown]. 

A treatment with less UVB irradiation episodes and/or inadequate intensity may 
fail to clear a psoriatic plaque, which eventually returns to the disease state 
once the treatment stops due to an insufficient loss in psoriatic stem cells. 
Figure~\ref{fig:recover}(b) and (c) show that after a 6-episode UVB irradiation 
treatment, the psoriatic stem cell and TA cell populations bounce back to the 
disease state after terminating the treatment. The total cell density drops to 
18600 mm$^{-2}$ at the end of the 6th episode with the normal and psoriatic stem 
cell densities as 247 and 1263 mm$^{-2}$, respectively. Interestingly, the 
end-treatment normal stem-cell count is slightly less than that from 7-episode 
treatment, implying that increased normal stem cell proliferation rate due to 
loss of TA cells well below the healthy level offsets the cell loss caused by 
apoptosis. Shortly after the treatment stopped, both psoriatic and normal stem 
cells and TA cells started slow increases, but later the density of psoriatic 
cells [Fig.~\ref{fig:recover}(b)] rapidly expands and outcompetes the normal 
cells [Fig.~\ref{fig:recover}(c)] that retreat from a maximum to the steady 
state at a lower level. During the entire course of the treatment the cytotoxic 
rate remains below the growth rate of psoriatic stem cells, giving no chance for 
the immune system to effectively reduce the psoriatic stem cell population (see 
Fig.~S3(d) in the Supplementary Materials). 

Histograms in Fig.~\ref{fig:recover}(d) and (e) show that a well-designed 
treatment can manage the psoriatic tissue to the quiescent state that is almost 
phenotypically indistinguishable from the healthy tissue in cell density 
distribution and the turnover time. Histologically, psoriasis causes thickening 
in the stratum corneum and the differentiated layer as well as an expanded 
proliferating compartment with more protruding rete 
ridges~\cite{iizuka2004psoriatic}. A dynamic model of rete ridge remodeling, 
simulation snapshots of the homeostatic psoriatic epidermis, tissue under 
treatment and recovered tissue can be found in the Supplementary Material 
(Fig.~S3-S4). Movies of simulations (normal and psoriatic tissues) are provided 
at URL: http://www.picb.ac.cn/stab/epidermal.html.

\section{Discussion}

We presented a hybrid model that simulates and visualizes spatiotemporal 
dynamics of the epidermal homeostasis. The model represents an efficient 
approach that separates the computation of cell kinetics from that of an 
agent-based cell migration. Compared to previous agent-based 
models~\cite{stekel1995computer,grabe2005multicellular,sun2007integrated,
adra2010development}, our population kinetics model describes cell 
proliferation, differentiation and cell death as empirical rate processes and 
can be simulated by integrating the governing ordinary differential equations 
(Table~\ref{tab:equations}) or the master equations by a kinetic Monte Carlo 
algorithm. The cell population kinetics model can be combined with a 
two-dimensional cell migration model to visualize dynamics of epidermis renewal 
and stratification. The model reproduces observed characteristics of the normal 
epidermis. Model analysis and simulations show that balancing cell production 
and cell loss in each subcompartment is critical to 
establishing and maintaining a proper epidermis homeostasis 
(Fig.~\ref{fig:odepopulation}).

The current model has some addressible limitations: (i) The cell population 
kinetics does not explicitly incorporate specific intracellular and 
extracellular factors that affect the dynamics and steady state of the 
epidermis 
homeostasis. However,  physiological and physical factors including age and UV 
irradiation as well as many commonly investigated signaling molecules can be 
coarsely coupled to model parameters such as proliferation and differentiation 
rate constants and morphology of keratinocytes and the epidermis. (ii) We 
neglected the effects by backconversions from TA cells to stem cells and from 
differentiated cells to TA cells by assuming their minimal impact. These 
processes could be worth a close examination as suggested in a recent 
theoretical study~\cite{chang2012uncontrolled} that showed rare backward 
transitions may cause catastrophic outcome such as a cancerous growth. (iii) 
Technically, as intensive modeling and computation is made possible by 
high-performance hardware~\cite{
christley2010integrative,Gord06102014}, our two-dimensional cross-sectional 
model can be extended to simulate a more realistic three-dimensional epidermis 
even though we expect that the qualitative results obtained from the 2D model 
remain valid in a 3D model.

As an important application, a non-trivial extension to the above model allows 
us to investigate the pathogenesis of psoriasis, an immune-mediated skin 
disorder. Genetic origins of psoriasis have been recently explored by an 
increasing number of genome-wide association studies that identified a multitude 
of psoriasis susceptibility 
loci~\cite{liu2008genome,nair2009genome,jordan2012rare,tsoi2012identification}. 
Many psoriasis-associated loci are connected to genes in the immune system 
(e.g., MHC class I molecules) and proteins expressed in keratinocytes, 
suggesting a complex nature of the disease~\cite{roberson2010psoriasis}. 
However, the mechanistic epidermis-immune system interactions implicated by 
these loci are yet to be resolved. 

In this study, we propose an alternative hypothesis of interactions between the 
immune system and keratinocytes, in which the disordered epidermis maintains a 
parallel homeostasis of both normal and psoriatic keratinocytes, derived from 
respective stem cell populations. We examine this hypothesis in an extended 
model and demonstrate that treatment by UVB irradiation with consecutive 
episodes can potentially manage the disease and achieve a remission of the 
psoriatic phenotype. 

Psoriasis has recently been studied by agent-based 
models~\cite{grabe2007simulating,weatherhead2011keratinocyte}. The model by 
Grabe and Neuber~\cite{grabe2007simulating} was able to generate the psoriasis 
phenotype of an increased cell density and a shortened epidermal turnover time, 
by adjusting the fractional time of TA cell proliferation. This parameter 
characterizes the amount of time for proliferation during a constant life time 
of a TA cell, which in our model is embedded in the TA cell self-proliferation rate 
constant $\gamma_2$. Increasing $\gamma_2$ and keeping symmetric division rate 
constant $k_{2s}$ unchanged (equivalent to keeping a constant TA cell life 
time) 
does increase TA cell population (Eq.~\ref{eq:scta}). To obtain a relative 
growth of the proliferating compartment, rate constants for cell differentiation 
must have relatively higher increases than $\gamma_2$, which was achieved by 
modulating Ca$^{2+}$ gradient in the Grabe and Neuber model. The model however 
did not propose possible management that can target the hypothesized mechanism. 
For example, it is not obvious how apoptosis induced by episodes of UVB 
irradiation can attain a remission via recovering the normal TA cell 
proliferating time. Weatherhead et al.~\cite{weatherhead2011keratinocyte} 
developed a model to simulate UVB-induced apoptosis in stem cells and TA cells, 
which was able to demonstrate a psoriasis remission after a few episodes of UVB 
irradiation. The model assumed a constant pool of stem cells that derived TA 
cells by asymmetric division and the UVB-induced apoptotic hyperproliferative 
stem cells were replaced with normal stem cells by symmetric divisions. This 
model assumption consequently led to a permanent reduction in cell density after 
each UVB irradiation treatment and therefore did not explain relapses of 
psoriatic phenotypes once an ineffective treatment ends or recurrence of the 
disorder.

Interactions between the immune system and keratinocytes considered in our model 
can serve as a conceptual basis for interpreting the pathogenesis of psoriasis. 
Especially, we showed that the act of the immune system cytotoxicity against 
psoriatic keratinocytes plays a pivotal role in the onset, remission and 
recurrence of the disease phenotype. The most common paradigm considers that 
faulty immune responses triggered by unknown self antigens or pathogens 
(introduced by injuries or trauma, known as Koebner phenomenon) produce 
cytokines and growth factors that promote keratinocyte hyperproliferation, 
immature differentiation and skin inflammation, establishing the psoriasis 
phenotype. Drugs that inhibit T cell activity and cytokine productions do 
improve psoriatic conditions~\cite{NEJMra0804595}. In contrast, our model 
demonstrates that a psoriasis lesion develops when the immune system is 
genuinely weakened or locally overwhelmed by a large population of psoriatic 
keratinocytes, staging up a chronic condition. This 
prediction may explain high occurrence and increasing severity of psoriasis in 
HIV-infected, especially late-stage AIDS patients with substantially compromised 
immune systems when CD4+ and naive CD8+ T cell counts substantially 
decrease~\cite{fife2007unraveling,morar2010hiv}. Furthermore, the onset age of 
psoriasis has been known to have two separate populations, type I (early onset 
in patients younger than 40 with a peak at age 20) and type II (late onset after 
age 40 with a peak at about 60)~\cite{queiro2014age}. Our model speculates that 
a vigorous immune system at a younger age can stimulate a strong 
hyperproliferation (a large $\rho_{sc}$) in a psoriatic epidermis and causes 
manifestation of plaque phenotypes. One the other hand, a weak immune system (a 
small $k_p$) at an older age can also have an equivalent effect.

The model prediction provides an alternative (a less explained) perspective of 
the pathogenesis of psoriasis, suggesting that psoriasis is a parallel 
epidermal homeostasis due to heterogeneity in stem cell clones, and that the immune 
system as a double-edged sword plays two essential however opposing roles: (i) 
Cytotoxic (CD8+) T cells recruited to the epidermis induce apoptosis in 
psoriatic stem cells, which is implicated by studies that demonstrated CD8+ T 
cell (especially, CD45RO+ memory subtype) count and cytotoxic proteins 
including perforin and granyme B substantially increase in psoriatic 
lesions~\cite{fife2007unraveling}; and (ii) Immune activities in the meantime 
promote progenitor keratinocytes proliferation by producing a multitude of 
cytokines and growth factors (TNF, IFN-$\alpha$, IFN-$\gamma$, IL-17, IL-22, 
IL-23, etc.). Balance between the two acts determines the outcome of the 
disease. Our model predicts that the psoriasis-susceptible tissue is a bimodal 
system and can switch between a non-symptom state and a phenotypical state, 
potentially explaining the recurring nature of the disorder and suggesting the 
feasibility of disease management by an effective treatment. This perspective 
and the model could be in general applicable to other autoimmune diseases in 
regenerative tissues. 

We showed in UVB phototherapy simulation that psoriasis plaque clearance can be 
attained by inducing strong apoptosis in keratinocytes. The model only 
considered UVB-induced apoptosis in keratinocytes. UVB irradiation may promote 
proliferation and differentiation in skin 
cells~\cite{lee2002acute,del2004ultraviolet} or may alter immune responses. 
These effects could be incorporated and be examined in the model. For example, 
as suggested in Eq.~(\ref{eq:ps2}), rebalance in self-proliferation 
$\gamma_{1,h}$ and symmetric division $k_{1s,h}$ and change in immune 
activities $f(\tilde{p}_{sc})$ may affect the population of psoriatic stem cells and thus 
the therapeutic outcome. Experiments did not demonstrate whether the UVB 
irradiation causes non-apoptotic cell death that was not reflected by an 
apoptosis marker~\cite{weatherhead2011keratinocyte}. Regardless the actual 
mechanism, induced cell deaths will result in a shift of the stem cell density 
from the disease state to the quiescent state across the phase boundary, an important parameter that determines the design of a phototherapy regimen, including irradiation dosage in each treatment episode, the number of episodes and the time interval between consecutive episodes. 

The model also suggests the limitation of UVB irradiation or similar treatments 
that attempt to achieve plaque remission by killing keratinocytes below a 
critical threshold. First, psoriasis plaques may recur under a 
temporally-weakened immune system or a transient burst of cell proliferation 
caused by conditions such as wound healing. Second, the psoriatic severity may 
be worsened under a genuinely weak immune system (with a low $K_p$ and/or a 
high threshold $K_a$), in which the disease phenotype persists and cannot be 
adequately reverted by simple reduction in psoriatic stem cells because the 
system does not possess a quiescent state [region I in Fig.~\ref{fig:phase}(b)]. In the latter case, other treatment options such as cytokine-targeting biologic and small-molecule drugs should be considered to at least shift the disorder to the bimodal region where phototherapy becomes effective. However, we note that 
phototherapy may be effective through alternatively mechanisms other than 
induction of apoptosis~\cite{racz2011effective,wong2013phototheratpy}, by which 
it alone may switch a plaque from persistent disorder (region I) to bimodal (region II) or symptomless 
(region III) as in Fig.~\ref{fig:phase}(b). 

Further insight from the model is that the bimodal psoriasis exhibits 
hysteresis, by which a mode of the epidermis, psoriatic or quiescent, once 
reached under a favorable condition, will tend to be relatively stable. For 
example, a symptomless epidermis as illustrated in Fig.~\ref{fig:phase}(b) may 
switch to the disease state when $K_p$ is reduced beyond its lower threshold 
because of weakening in the immune system. However, restrengthening the immune 
system to revert the disease state back to the quiescent mode requires elevating 
$K_p$ in the model beyond the upper threshold, implying that the disease is 
resistant to mild natural or induced perturbations. 

One definitive experimental test of our model hypothesis should aim at 
simultaneously tracing clones of psoriatic and normal stem cell lineages. 
Because of recent advancement in techniques, studies of epidermal stem cells 
become central to the understanding of the epidermal homeostasis. Especially, 
the emerging powerful {\em in vivo} lineage tracing technology allows dynamic 
monitoring of stem cells and progenitors clone formation and differentiation and 
can achieve multicolor simultaneous tracing of separate clones~\cite{alcolea2013tracking}. The technique has already been pioneered to study progenitor fates in normal skin tissue~\cite{clayton2007single} and benign 
epidermal tumors~\cite{driessens2012defining} and can be potentially applicable 
to identify the coexistence of normal and psoriatic progenitors in psoriasis. 
One potential challenge is that in either quiescent or disease state one type of 
stem cells is in a small population (about a 1:20 ratio, according to Table~\ref{tab:distro}), making it difficult for differential labeling. Alternatively, an experiment may tend to identify and isolate the two competing effects by the immune system on the epidermis: (1) apoptosis in epidermal progenitors by T-cell cytotoxic activities 
and (2) hyperproliferation due to elevated cytokines and growth factors.

\appendix
\section{Analytical details of the psoriasis model}
\renewcommand{\theequation}{A\arabic{equation}}
\setcounter{equation}{0}

\begin{figure}[t]
\centering
\includegraphics[scale=0.33]{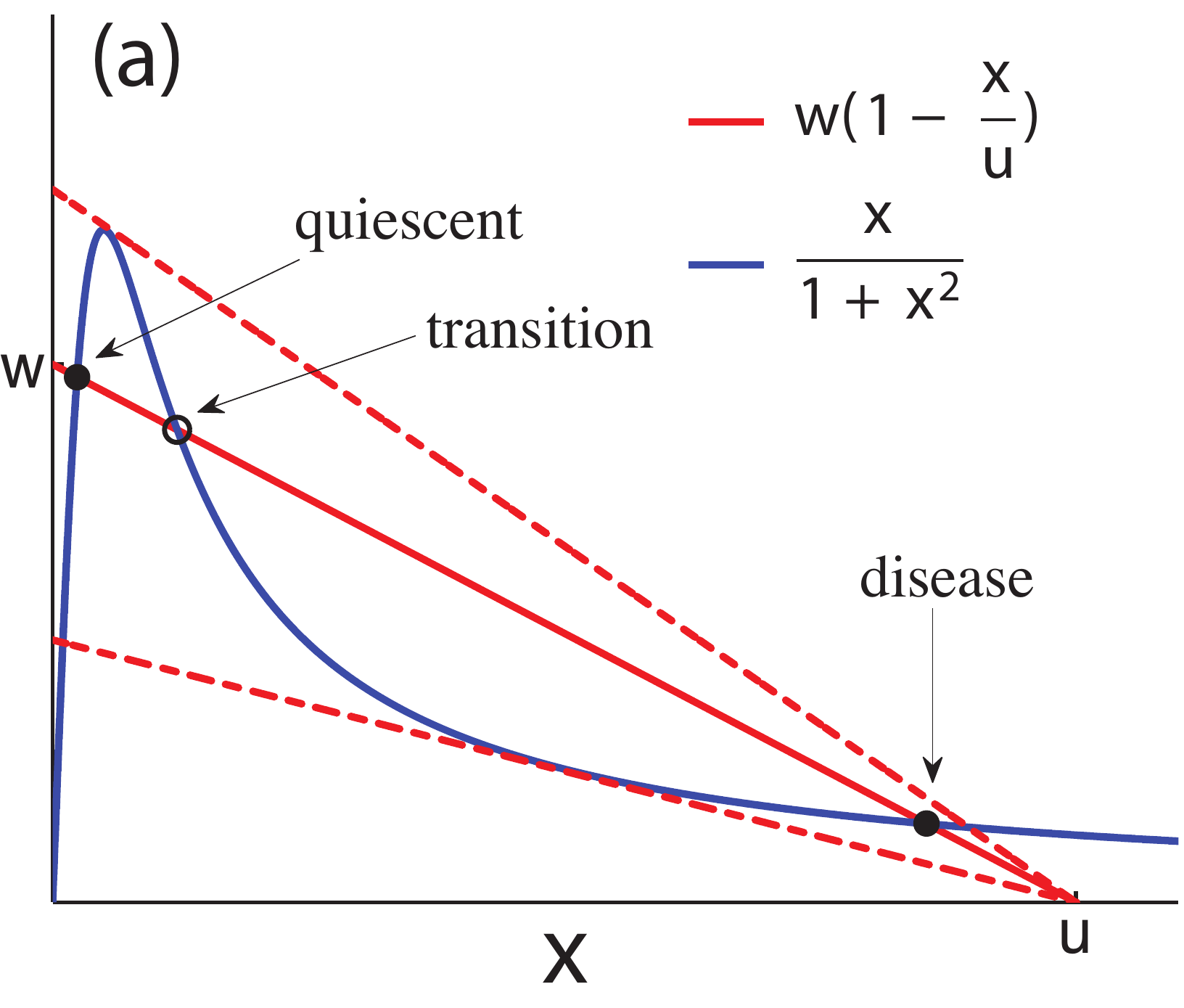}
\includegraphics[scale=0.33]{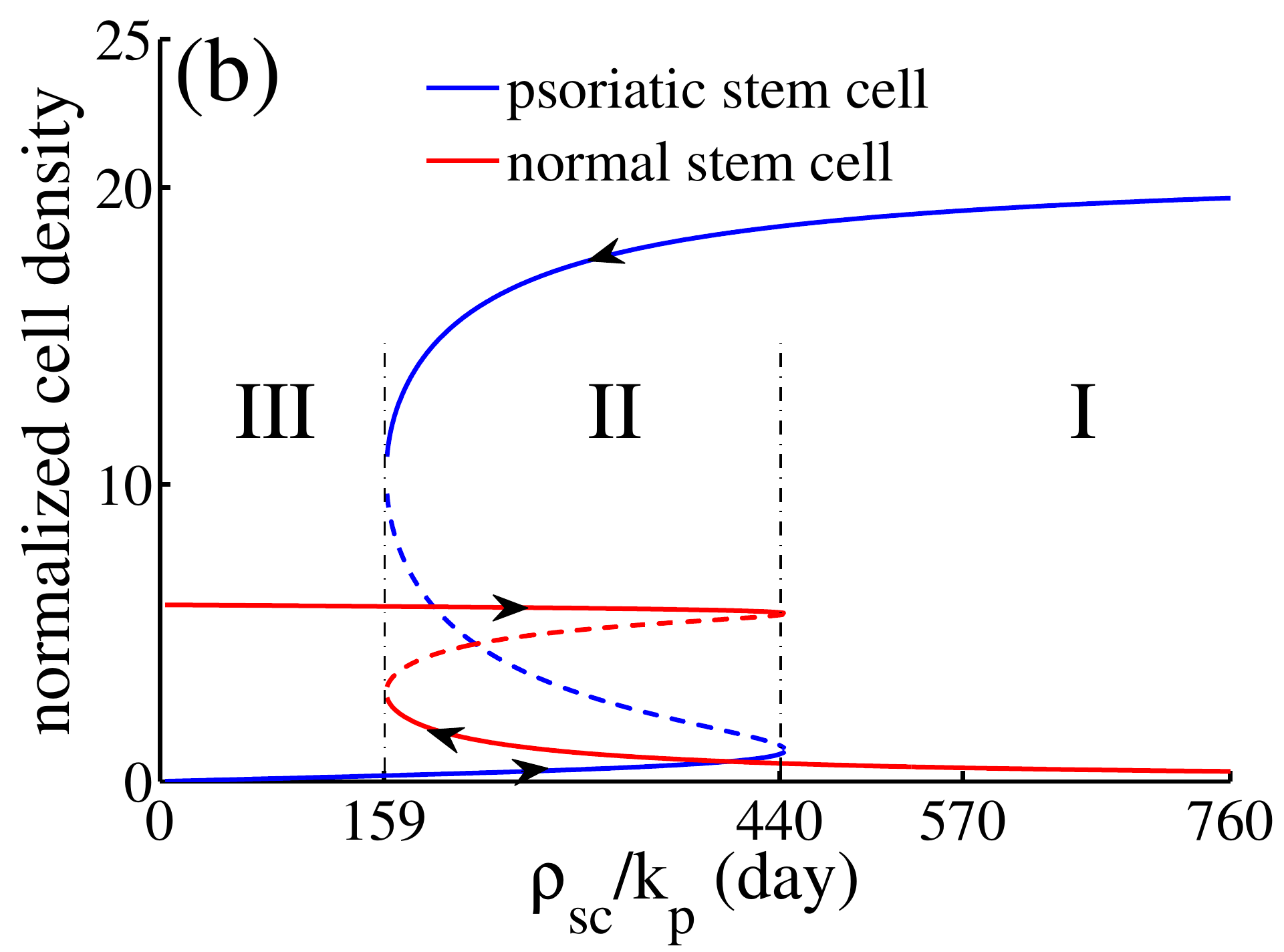}
\includegraphics[scale=0.33]{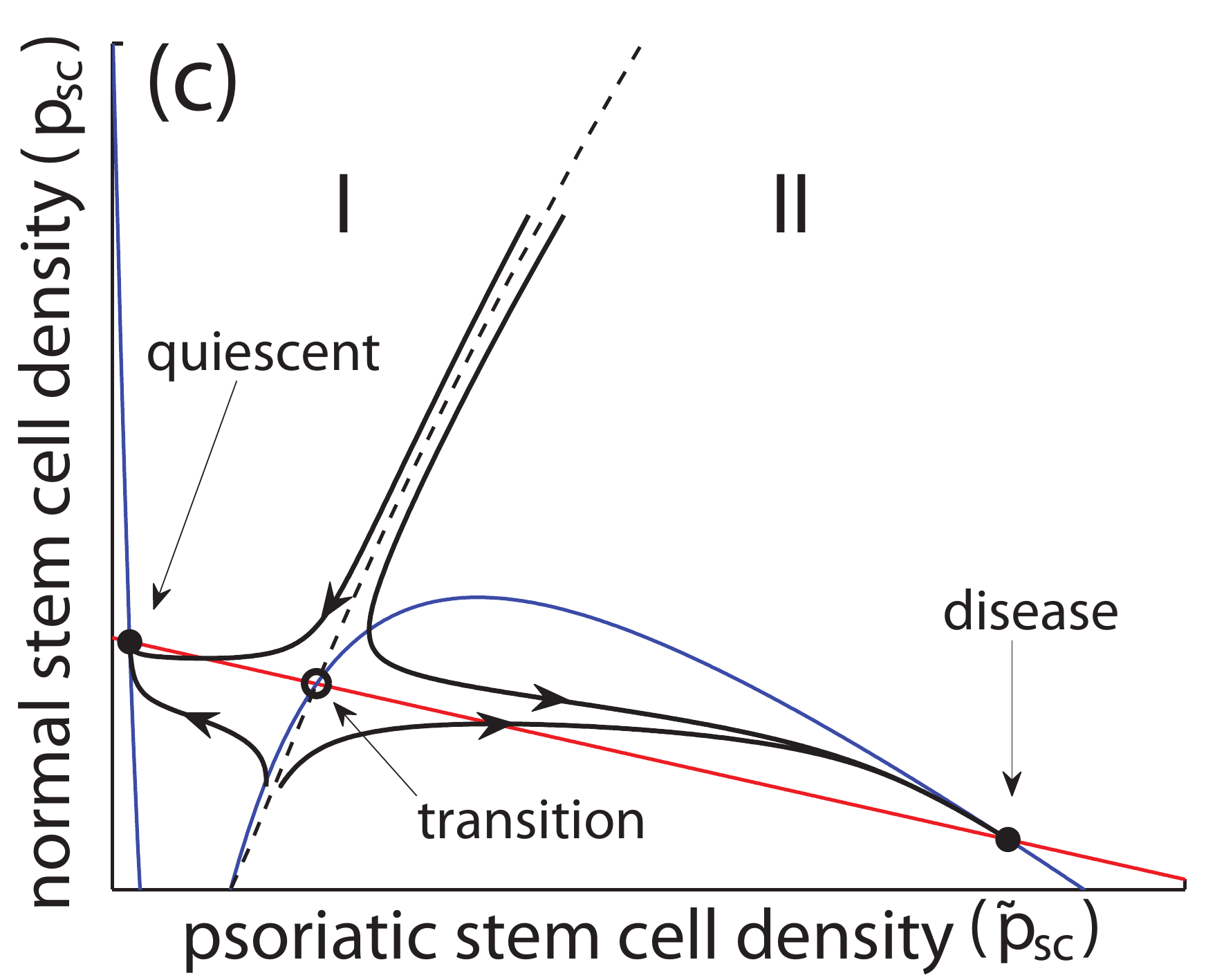}
\caption{\label{fig:phase} {\bf Steady states of the psoriasis model.}  (a) Intersections of the two functions give the steady states (Eq.~\ref{eq:x}) and can be modulated with $w$ by changing $K_p$ within the range of two bifurcation points (indicated by dashed lines). The ``quiescent" and ``disease" states (filled circles) are stable steady states. The ``transition" state (empty circle) is unstable (i.e., arbitrarily small perturbations deviate the system away from it). (b) Bifurcation diagram under varying ration $\rho_{sc}/k_p$ with three regions: (I) persistent disorder, (II) bimodal, and (III) symptomless. (c) Trajectories from basins of attraction: I  and II, separated by the boundary (dashed line), converging to the quiescent and disease states, respectively. Blue and red curves are nullclines (Eqs.~\ref{eq:x} and \ref{eq:y}). Parameter values used for the simulation are listed in Table~\ref{tab:parameter}.}
\end{figure}

\noindent {\em Psoriasis as a bimodal switch ---} To illustrate the principle 
and simplify the analysis, we only present a two-dimensional model that 
describes dynamic interactions between normal and psoriatic stem cells. The 
complete model that produced the results (Fig.\ref{fig:recover}) is more complex 
and of a higher dimension because of the feedback regulation by TA cell 
population on the normal stem cell proliferation (Eq.~\ref{eq:scrateps}), 
requiring analyzing the dynamics of normal and psoriatic TA cells together. 
Again, for convenience in analysis, we neglect the influences of backconversions 
and apoptosis. This simplification allows us to show insights of the model 
behaviors without much mathematical complication. We work with normalized 
quantities:
\begin{equation}
x=\tilde{p}_{sc}/K_a, \ \ y=p_{sc}/K_a, \ \  y^{\rm max}=p_{sc}^{\rm max}/K_a, \ 
\  k_p=K_p/K_a \notag \ .
\end{equation}
Ratios between the normal stem-cell proliferation rate constants remain constant 
despite their dynamic modulations by the TA cell population 
(Eq.~\ref{eq:scrate}), i.e., $k_{1s}/\gamma_1=k_{1s,h}/\gamma_{1,h}$. Ignoring 
the trivial steady states of zero density ($x=0$, and/or $y=0$), based on 
Eqs.~(\ref{eq:ps1}-\ref{eq:ps3}) we have the steady-state equations for stem 
cell densities as
\begin{equation}\label{eq:x}
w\left(1-\frac{x}{u}\right)-\frac{x}{1+x^2} =0 \ ,
\end{equation}
\begin{equation}\label{eq:y}
y= \frac{u-x}{\lambda} \ ,
\end{equation}
where
\begin{eqnarray}
w&=& \frac{\rho_{sc}}{k_p}(1-1/\lambda)(\gamma_{1,h}-k_{1s,h})\notag \ , \\
u&=& \lambda y^{\rm max}(1-k_{1s,h}/\gamma_{1,h}) \notag\ .
\end{eqnarray}
Eq.~(\ref{eq:x}) can solve for one or three steady-state psoriatic stem cell 
densities, only depending on choices of two parameters, $w$ and $u$, which is 
illustrated in Fig.~\ref{fig:phase}(a).

To study plaque remission and relapse, we are interested in the region where the 
system has two stable steady states, ``quiescent" and ``disease", and one 
unstable ``transition" steady state. This scenario is illustrated in 
Fig.~\ref{fig:phase}(a) with three intersections between the line $w(1-x/u)$ and 
the nonparametric curve $x/(1+x^2)$. Determination of the stability of these 
fixed points requires analysis similar to the spruce budworm model by Ludwig et 
al.~\cite{ludwig1978qualitative}. Figure~\ref{fig:phase}(a) shows that the 
``quiescent" state is more sensitive to $w$, whereas the ``disease" state is 
determined by both $u$ and $w$. This observation provides a guideline to 
parameterize the model to attain appropriate psoriatic phenotypes by identifying 
disease-related parameters $\rho_{sc}$, $\lambda$ and $k_p$.  The ratio 
$\rho_{sc}/k_p$ emerges as the key parameter that characterizes the interplay 
between proliferation of keratinocytes and activity of the immune system 
(magnitude $K_p$ and threshold 
$K_a$), playing a critical role in the interpretation of the pathogenesis of 
psoriasis. 

Bifurcation diagrams in Fig.~\ref{fig:phase}(b) of steady-state stem cell 
densities shows three classes of behaviors as $\rho_{sc}/k_p$ varies. A small 
population of psoriatic stem cells survive in the ``quiescent" state 
($\rho_{sc}/k_p<$159 day) under a vigorous cytotoxic activity (a large $k_p$, 
i.e., a low threshold $K_a$ and/or a high magnitude $K_p$) and/or a moderate 
stem-cell proliferation (a small $\rho_{sc}$). In contrast,  a weak immune 
system (a small $k_p$, i.e., a high threshold $K_a$ and/or a low magnitude 
$K_p$) cannot adequately counterbalance the psoriatic hyperproliferation (a 
large $\rho_{sc}$) and thus the epidermis assumes a persistent ``disease" state 
($\rho_{sc}/k_p>$440 day).  A bimodal system with an intermediate 
$\rho_{sc}/k_p$ has a potential to switch between the quiescent and disease 
modes when conditions change. Figure~\ref{fig:phase}(c) shows a two-dimensional 
phase plane of normal and psoriatic stem-cell densities. Starting from initial 
stem cell densities, located inside 
region I or II, a temporal trajectory will be attracted to the quiescent or the 
disease state, respectively. Those starting on the boundary of the two regions 
will in theory converge to the transition state, which is unsustainable because 
slight perturbations will dislocate a trajectory into either region I or II.

\noindent {\em Cell density and turnover time ---} Below, we provide equations 
for homeostatic cell densities and the turnover time in the psoriatic model. As 
shorthands, we define
\begin{equation}
g=\frac{k_{1a}+2k_{1s}}{k_{2s}-\gamma_2}, \ \ \ 
g_h=\frac{k_{1a,h}+2k_{1s,h}}{k_{2s}-\gamma_2} \notag \ ,
\end{equation}
The ratio $\xi=g/g_h$ is a function of the steady-state TA cell population 
($p_{ta}+\tilde{p}_{ta}$) by Eq.~(\ref{eq:scrateps}). The total steady-state 
cell density is given as
\begin{equation}
p_{\rm tot}= p_{sc}\left(1+\xi g_hW\right)+ 
\tilde{p}_{sc}\left(1+\frac{\rho_{sc}}{\rho_{ta}}g_h\widetilde{W}\right) 
\notag \ ,
\end{equation}
where
\begin{eqnarray}
W&=&1+(k_{2a}+2k_{2s})\left(\frac{1}{k_3}+\frac{1}{k_4}+\frac{1}{k_5}+\frac{1}{
\alpha}\right)\notag  \ ,\\
\widetilde{W}&=&1+(k_{2a}+2k_{2s})\left(\frac{1}{\rho_{tr}k_3}+\frac{1}{\rho_{tr
}k_4}+\frac{1}{\rho_{de}\alpha}\right)\rho_{ta} \notag \ .
\end{eqnarray}
Notice the lack of GC layer in the psoriatic cell population. The proliferative 
and differentiated cell populations are
\begin{equation}
p_{\rm prolif}=p_{sc}(1+\xi 
g_h)+\left(1+\frac{\rho_{sc}}{\rho_{ta}}g_h\right)\widetilde {p}_{sc} \notag
\end{equation}
\begin{eqnarray}
p_{\rm 
diff}&=&g_h(k_{2a}+2k_{2s})\left(p_{sc}\xi+\frac{\rho_{sc}}{\rho_{tr}}\widetilde
{p}_{sc}\right)\left(\frac{1}{k_3}+\frac{1}{k_4}\right) \notag \\
& &+p_{sc}\xi g_n\frac{k_{2a}+2k_{2s}}{k_5} \notag \ .
\end{eqnarray}
Corneocytes are not included in the differentiated compartment. By comparison, 
the cell densities in the healthy tissue are
\begin{equation}
p_{\rm prolif,h}=p_{sc,h}(1+g_h) \notag \ ,
\end{equation}
\begin{equation}
p_{\rm 
diff,h}=g_h(k_{2a}+2k_{2s})p_{sc,h}\left(\frac{1}{k_3}+\frac{1}{k_4}+\frac{1}{
k_5}\right) \notag \ .
\end{equation}
We can calculate the growth of the proliferative compartment relative to that of 
the nonproliferative compartment as
\begin{eqnarray}
\frac{p_{\rm prolif}/p_{\rm prolif,h}}{p_{\rm diff}/p_{\rm 
diff,h}}&=&\frac{p_{sc}(1+\xi 
g_h)+\left(1+\frac{\rho_{sc}}{\rho_{ta}}g_h\right)\tilde{p}_{sc}}{(1+g_h)}
\notag \\
& \times 
&\frac{\frac{1}{k_3}+\frac{1}{k_4}+\frac{1}{k_5}}{\left(p_{sc}\xi+\frac{\rho_{sc
}}{\rho_{tr}}\tilde{p}_{sc}\right)\left(\frac{1}{k_3}+\frac{1}{k_4}\right)+p_{sc
}\xi\frac{1}{k_5}} \notag \\ &>&\frac{p_{sc}(1+\xi 
g_h)+\left(1+\frac{\rho_{sc}}{\rho_{ta}}g_h\right)\tilde{p}_{sc}}{
(1+g_h)\left(p_{sc}\xi+\frac{\rho_{sc}}{\rho_{tr}}\tilde{p}_{sc}\right)} 
\notag \ .
\end{eqnarray}
A relative increase in the proliferative population requires the above ratio to 
be greater than 1.  In psoriasis, $\xi<1$, reflecting a decreased proliferation 
rate in the cohabitating normal stem cell population due to a much elevated TA 
cell population. It is unclear whether stem cells and TA cells have differential 
increases in proliferation rates. For the lack of information, we assume 
$\rho_{sc}/\rho_{ta}\approx 1$. Therefore, $\rho_{tr}>\rho_{sc}$ becomes a 
sufficient (but not a necessary) condition to produce a relative increase in the 
proliferative compartment, requiring that the fold increase of transit rate in 
the differentiated compartment is higher than the fold increase of division 
rates in the proliferative compartment. This result agrees with an earlier 
prediction from Heenen et al.~\cite{heenen1987psoriasis}, showing that a 
relative growth in the proliferative compartment requires keratinocyte 
hyperproliferation and  in the meantime an increased transit time in the 
differentiated compartment. A proportional increase in both TA-cell 
proliferation rate and transit rate ($\rho_{sc}=\rho_{tr}$) does not cause 
relatively differential growth of the two compartments unless difference exists 
between increases of proliferating rates of stem cells and TA cells, 
$\rho_{sc}\ne\rho_{ta}$. Fast migration of differentiated keratinocytes also 
offsets overgrowth in the proliferative compartment. An important insight from 
the model is that increasing keratinocyte population in psoriasis is not merely 
caused by elevated proliferation rates but also requires the increase in growth 
capacity in the progenitor repertoire. Observed morphological remodeling of the 
psoriatic epidermis suggests increased growth niches for stem 
cells~\cite{iizuka2004psoriatic}.

Similarly to Eq.~(\ref{eq:to}), the epidermal turnover time can be approximated 
as the ratio of the total cell density to the rate of desquamation:
\begin{eqnarray}
\tilde{\tau}_{\rm tot} &\approx&\frac{p_{\rm 
tot}}{\alpha(p_{cc}+\rho_{de}\tilde{p}_{cc})} \notag\\
&=& 
\left(\frac{p_{sc}+\tilde{p}_{sc}}{p_{sc}\xi+\rho_{sc}\tilde{p}_{sc}}\frac{1}{
g_h}+\frac{p_{sc}\xi+\frac{\rho_{sc}}{ 
\rho_{ta}}\tilde{p}_{sc}}{p_{sc}\xi+\rho_{sc}\tilde{p}_{sc}}\right)\frac{1}{k_{
2a}+2k_{2s}} \notag \\
&&+\frac{p_{sc}\xi+\frac{\rho_{sc}}{\rho_{tr}}\tilde{p}_{sc}}{p_{sc}\xi+\rho_{sc
} \tilde{p}_{sc}}\left(\frac{1}{k_3}+\frac{1}{k_4}\right) \notag\\
&&+\frac{p_{sc}\xi}{p_{sc}\xi+\tilde{p}_{sc}\rho_{sc}}\frac{1}{k_5}+\frac{p_{sc}
\xi+\frac{\rho_{sc}}{\rho_{de}}\tilde{p}_{sc}}{p_{sc}\xi+\rho_{sc}\tilde{p}_{sc}
}\frac{1}{\alpha}  \notag \ ,
\end{eqnarray}
which can be analyzed for the two psoriasis modes (quiescent or disease). At the 
quiescent  mode, $\xi\approx 1$ and $p_{sc}\gg \tilde{p}_{sc}$, the turnover 
time is approximated by
\begin{equation}
\tilde{\tau}_{\rm tot}\approx 
\left(\frac{k_{2s}-\gamma_2}{k_{1a}+2k_{1s}}+1\right)\frac{1}{k_{2a}+2k_{2s}}
+\frac{1}{k_3}+\frac{1}{k_4}+\frac{1}{k_5}+\frac{1}{\alpha} \notag \ ,
\end{equation}
converging back to $\tau_{\rm tot}$. At the disease mode, $\xi<1$ and $p_{sc}\ll 
\tilde{p}_{sc}$. The turnover time is close to:
\begin{equation}
\tilde{\tau}_{\rm tot}\approx 
\left(\frac{1}{\rho_{sc}g_h}+\frac{1}{\rho_{ta}}\right)\frac{1}{k_{2a}+2k_{2s}} 
+\frac{1}{\rho_{tr}}\left(\frac{1}{k_3}+\frac{1}{k_4}\right)+\frac{1}{\rho_{de}}
\frac{1}{\alpha}\notag \ .
\end{equation} 
Both cases do not account for the relatively much smaller cell loss by the 
cytotoxic effect from the immune system. Compared to the turnover time of the 
healthy tissue $\tau_{\rm tot}$ (Eq.~\ref{eq:to}), $\tilde{\tau}_{\rm tot}$ of 
the psoriatic tissue is clearly shortened at each stage from the proliferative 
compartment (the first term) to differentiated compartment (the second term) and 
to the corneum (the third term). The turnover time of the granular layer is 
ignorable because of the minimal normal cell population.

\section*{Acknowledgments}
This work was supported by LVMH Recherche through Sprim Inc. and institutional 
fund to JY. We thank Weiren Cui, Eric Perrier, Micha$\ddot{\mathrm{e}}$l 
Shleifer, Gallic Beauchef and Delphine de Queral for helpful discussions.

%\bibliographystyle{unsrt}
%\bibliography{reference}

\clearpage
\renewcommand{\thefigure}{S\arabic{figure}}
\setcounter{figure}{0}

\renewcommand{\theequation}{S\arabic{equation}}
\setcounter{equation}{0}

\section*{Supplementary Materials}

\small
\subsubsection*{Computation of population kinetics model}

\begin{figure}[t]
\centering
\includegraphics[scale=0.3]{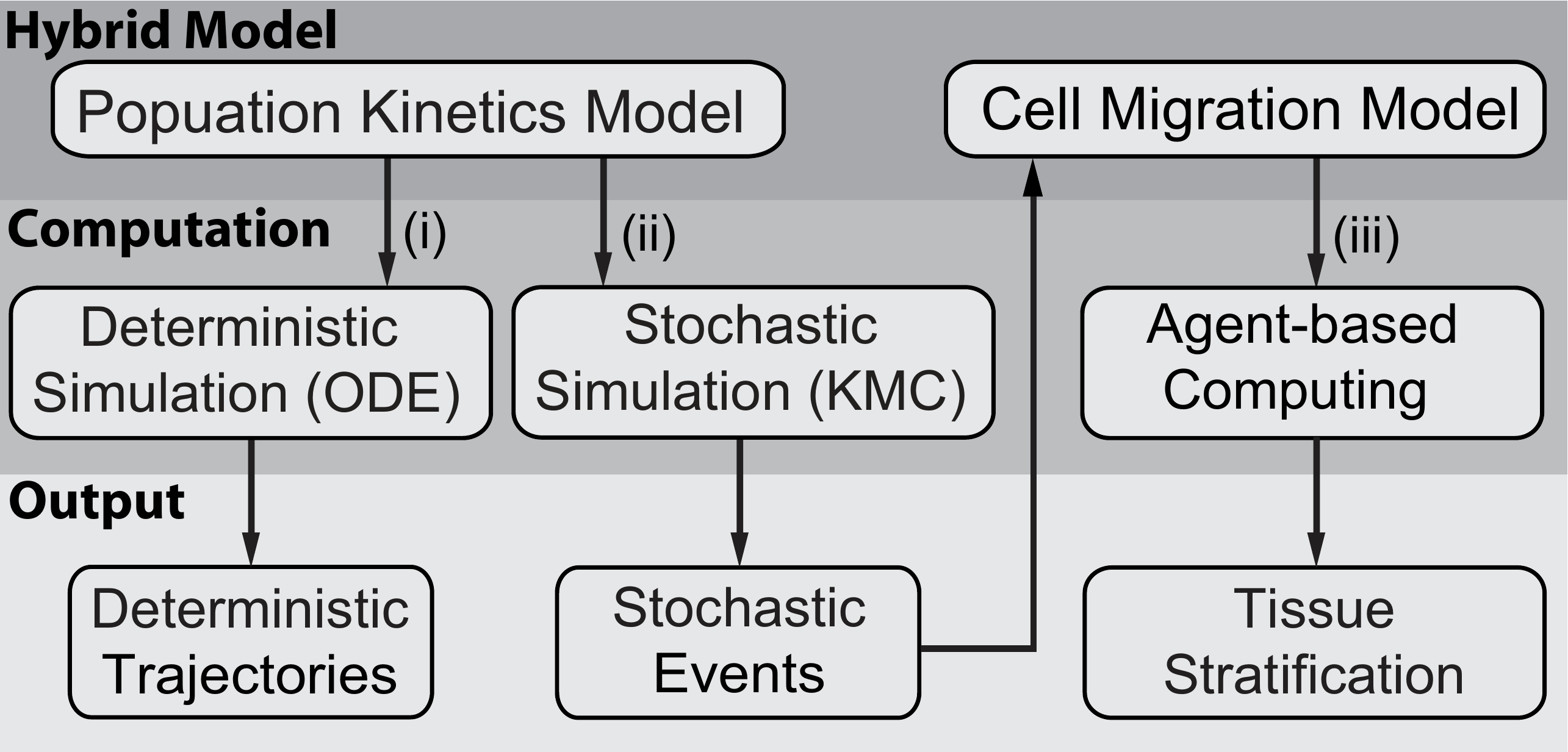}
\caption{\small {\bf Schemes of model computation.} Kinetics of the central 
transition pathway of the epidermis renewal can be computed by one of the two 
numerical methods: (i) integrating the system of ODEs, or (ii) stochastic 
simulation algorithm (SSA). As a third option (iii), the spatio-temporal 
dynamics of the epidermis renewal can be computed and visualized by 
incorporating cell events generated by the stochastic simulation into the cell 
migration model.}\label{fig:diagram}
\end{figure}

The central transition pathway of the epidermis renewal can be computed in two 
alternative ways (Fig.~\ref{fig:diagram}) to obtain the {\em deterministic} or 
{\em stochastic dynamics} of the system. The mathematical equations [the 
ordinary differential equations (ODEs) or the master equations] of the system 
for the normal tissue can be found in Table 1 of the paper. 

\noindent {\em Deterministic simulation ---} The deterministic model describes 
cell populations by a set of ODEs with the cell densities as the state variables 
according to the rate processes in the central transition pathway. Numerically 
integrating the ODEs obtains trajectories of the cell density distribution over 
all individual cell types.

\noindent {\em Stochastic simulation ---} In this case, rate processes in the 
central transition pathway are sampled by the Gillespie's 
algorithm~\cite{gillespie2007stochasticsp} to generate probabilistic cellular 
events sequentially in time. An individual cell process is chosen by a 
probability proportional to the rate of the process, and the occurrence time 
interval is sampled as in a Poissonian process with a mean rate equal to the 
accumulated rate of all processes in the system. Time series of cell events 
generated by the stochastic simulation are fed to the agent-based cell migration 
model to simulate and visualize the spatiotemporal organization of keratinocytes 
in the epidermis.

\subsubsection*{Computation of agent-based cell migration model}

Simulation of the cell migration model is coupled with the stochastic 
simulation of the cell population kinetics model described in the previous 
section. Because the cell population kinetics is treated independently from the 
cell migration in our model, the stochastic simulation of population dynamics 
can be completed before simulating the cell migration, and the generated 
cellular events of proliferation, differentiation, apoptosis and desquamation 
can be stored and later embedded into the time series of migration events. \\ 

\noindent {\em Repulsive force ---} The amount of repulsive force ${\bf 
F}_{ij}^r$ is assumed to be proportional to the overlapping area $S_{ij}$ and 
the direction of the force is along the line connecting the cell centers of 
cells $i$ and $j$. At time $t$, the two orthogonal components ${\bf 
F}_{ij}^r(t)\equiv [F_{ij}^{r,x}(t)\, , F_{ij}^{r,y}(t)]$ is given as 
\begin{equation}\label{eq:ol}
F_{ij}^{r,x}(t)=kS_{ij}(t)\cos\theta_{ij}, \  \text{and} \ 
F_{ij}^{r,y}(t)=kS_{ij}(t)\sin\theta_{ij} \ ,
\end{equation}
where $\theta_{ij}$ is the angle of vector ${\bf F}_{ij}^r$ and $k$ is a 
constant assumed identical for all cell types.

\begin{figure}[b]
\centering
\includegraphics[scale=0.3]{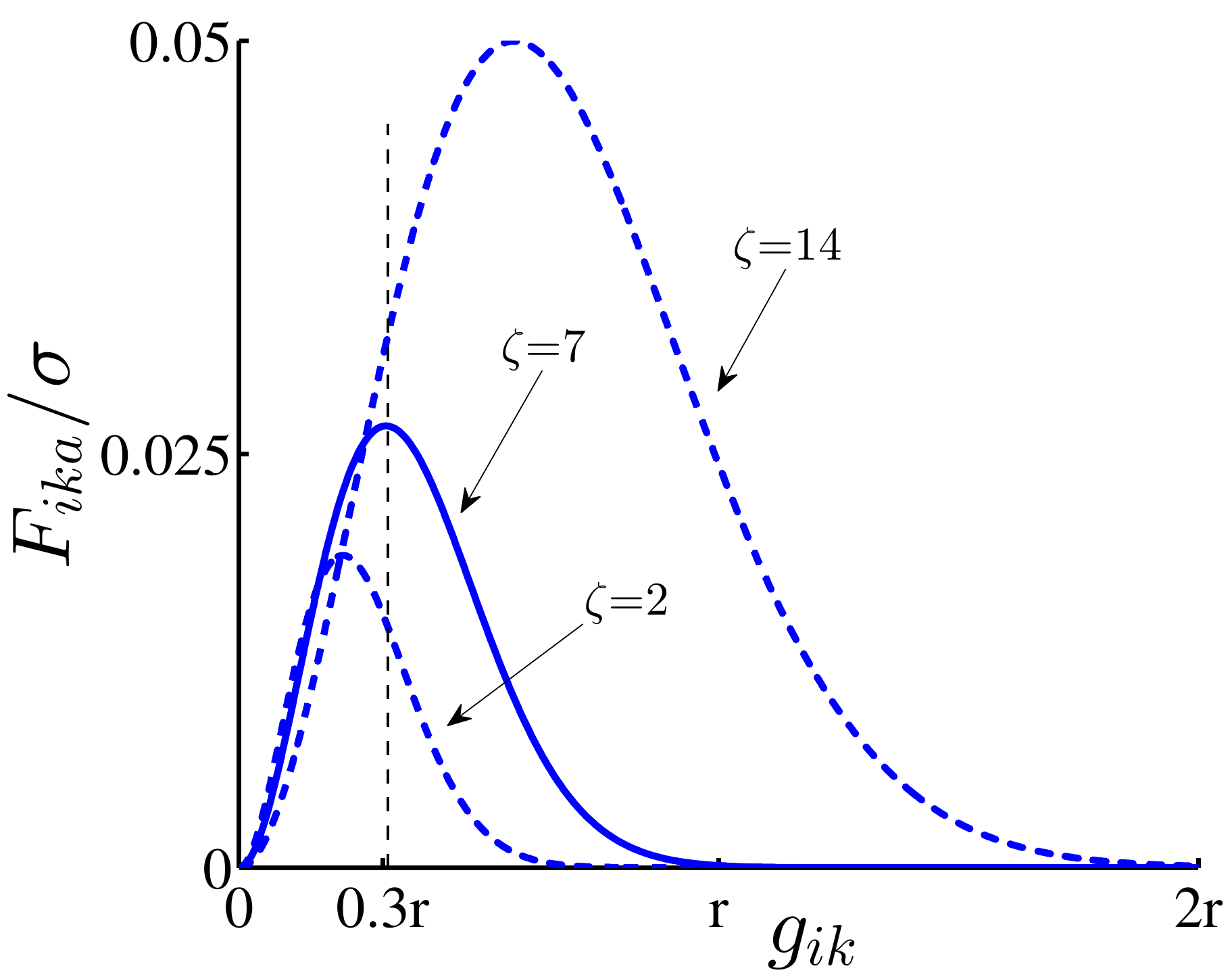}
\caption[{\bf Adhesive force}]{\small{\bf Adhesive force}. The normalized 
adhesive force ($F_{ik}^a/\sigma$) related the cell-cell distance under 
different $\zeta$ values. The adhesive force decreases to zero at $\zeta=7$ 
when 
the cell-cell distance exceeds the cell radius $r$. \label{fig:adhesion}} 
\end{figure}

In theory, the overlapping area between two ellipsoids can be precisely 
calculated by a numerical integration. To increase computational efficiency,  
$S_{ij}$ (note that $S_{ij}=S_{ji}$) is however approximated by the overlap 
between the two cells as if they are rectangles with widths and heights 
identical to the corresponding ellipsoids. An overlap happens between cells $i$ 
and $j$ if the following two conditions hold: 
\begin{equation}
|x_i-x_j|<\frac{w_i+w_j}{2}, \ \ \ \text{and} \ |y_i-y_j|<\frac{h_i+h_j}{2} \ ,
\end{equation}
where $h_i$ and $w_i$ are height and width of cell $i$, respectively. The width 
and height, $w_{ij}$ and $h_{ij}$, of the overlap rectangle can then be 
calculated as 
\begin{eqnarray}
w_{ij}&=& \min\{\frac{w_i+w_j}{2}-|x_i-x_j|, \ w_i,\ w_j\} \\
h_{ij}&=& \min\{\frac{h_i+h_j}{2}-|y_i-y_j|, \ h_i,\ h_j\} \ ,
\end{eqnarray}
and the overlap area is approximated as $S_{ij}\approx w_{ij}h_{ij}$. \\

\begin{figure*}[t]
\centering
\includegraphics[scale=0.35]{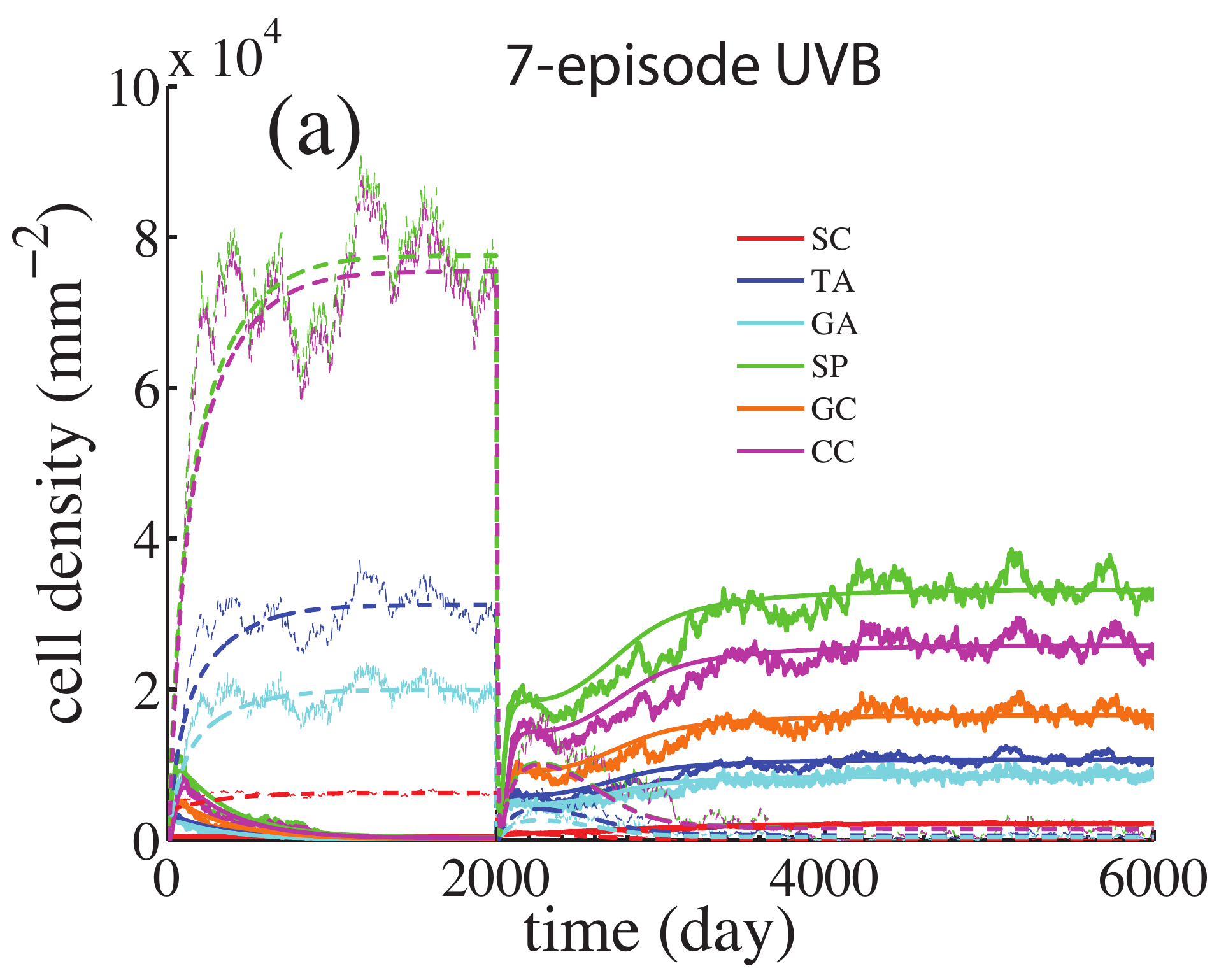}
\includegraphics[scale=0.35]{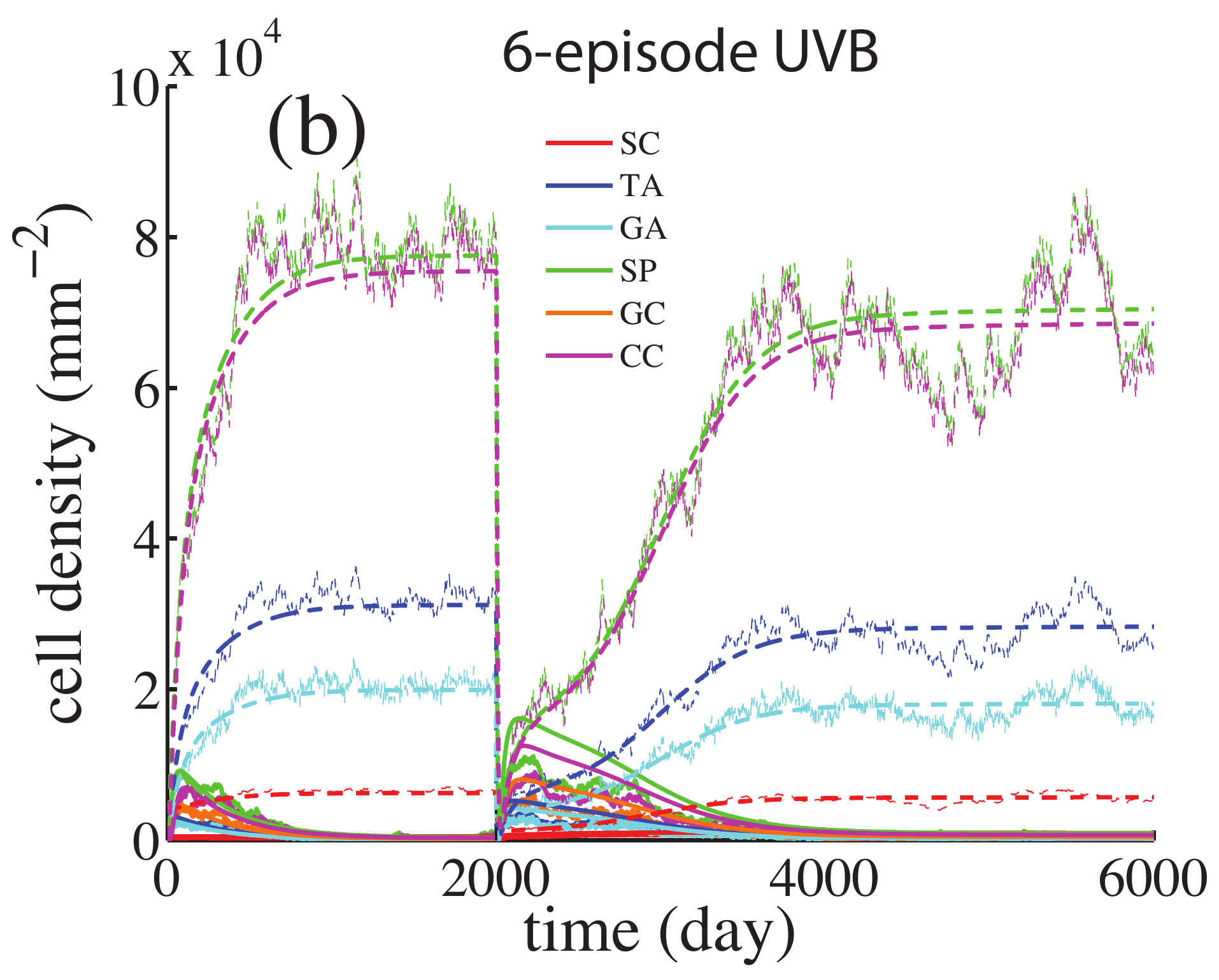}
\includegraphics[scale=0.35]{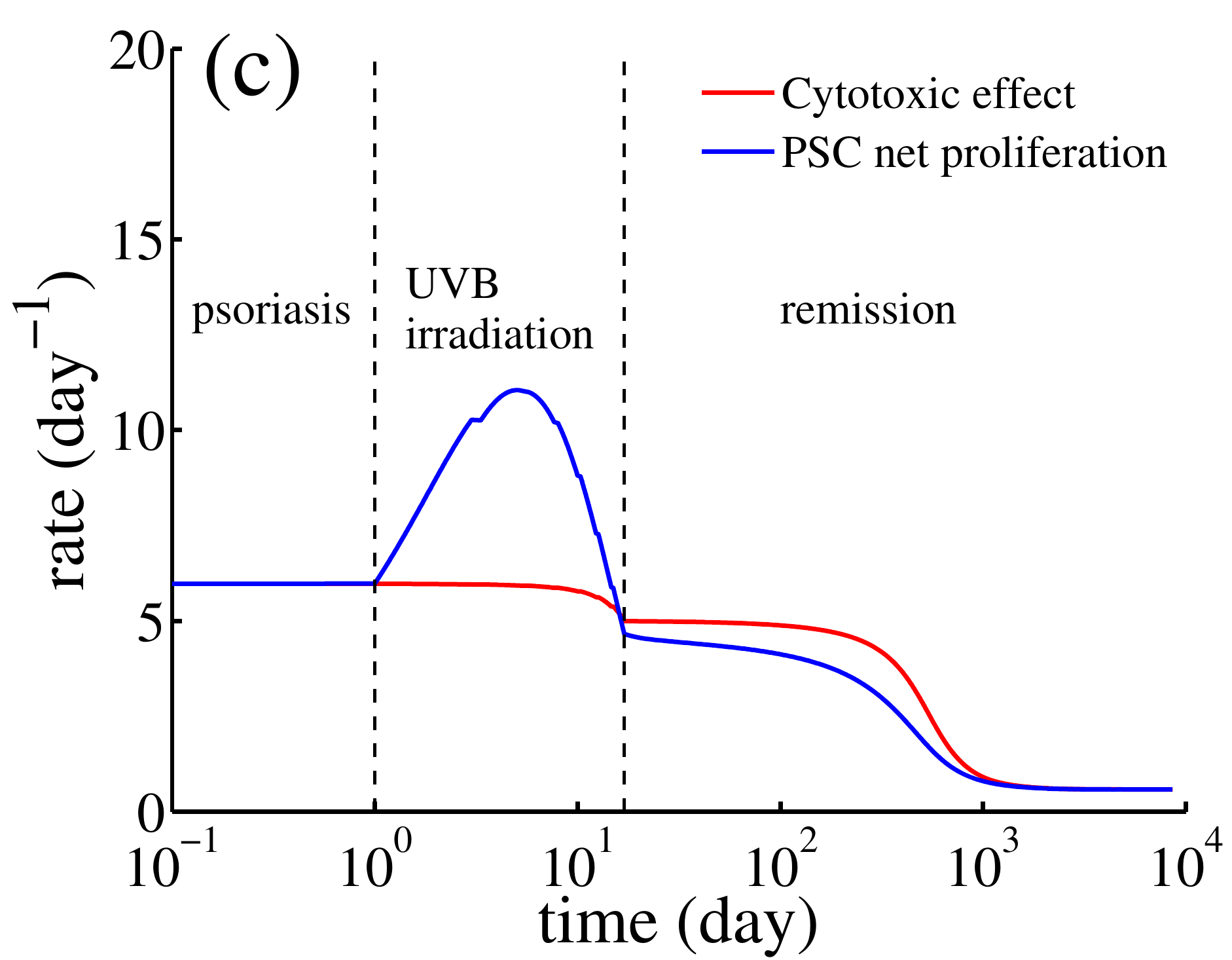}
\includegraphics[scale=0.35]{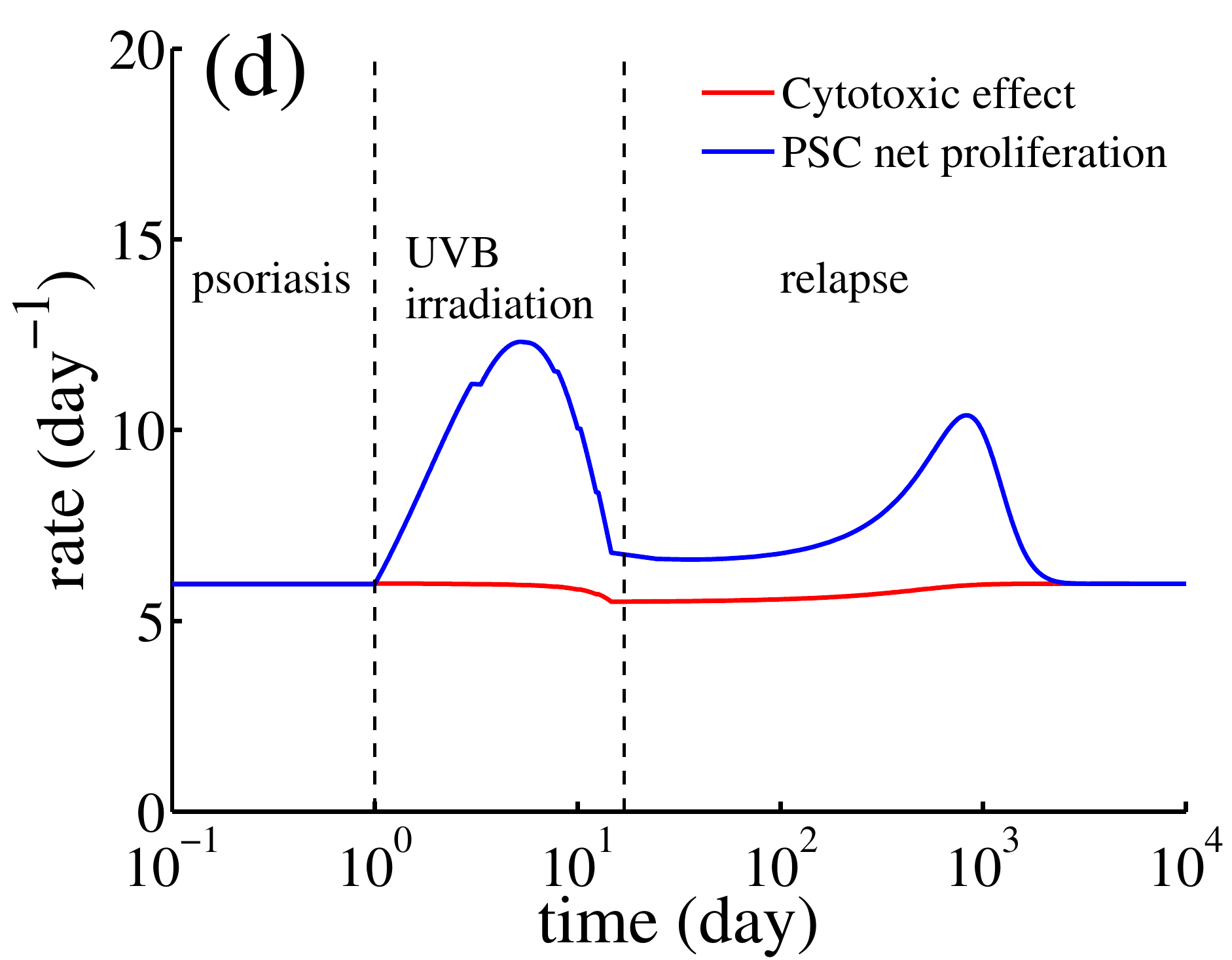}
\caption{\small {\bf Psoriasis phototherapy by multi-episode UVB irradiation. 
\label{fig:s3}} (a) 7-episode and (b) 6-episode UVB irradiation treatments. The 
model simulated an area of length 5 mm $\times$ width 0.01 mm =0.05 mm$^2$ 
(i.e., a single-cell sheet of the epidermis). Simulations started with a mixed 
population of normal (solid) and psoriatic (dashed) stem cells (about a 20:200 
ratio).  Temporal evolution of the cytotoxic killing rate and psoriatic stem 
cell (PSC) net proliferation rate (selfproliferation rate - symmetric division 
rate) in (c) 7-episode and (d) 6-episode UVB irradiation. Note that compared to 
(a) and (b), time axes are shifted in plots (c) and (d) to remove the initial 
transient dynamics to psoriatic homeostasis. The UVB treatments follow the same 
protocol described in the main text (Fig. 6). Parameter values used in the 
simulations are listed in the main text (Table 3).}
\end{figure*}

\noindent {\em Adhesive force ---} We adapt the treatment of cell-cell adhesive 
force from Palsson model~\cite{palsson2008sp} and Li et 
al.~\cite{li2013skinsp}. The 
adhesive force between two adjacent cells $i$ and $j$, ${\bf F}_{ij}^a$, is a 
biphasic function of the cell-cell distance and the direction of the force is 
along with the line connecting the cell centers. We have the following equation 
for ${\bf F}_{ij}^a$ (see Fig.~\ref{fig:adhesion} for plots of force-distance 
relationship):
\begin{equation}\label{eq:af}
F_{ij}^a=-\sigma\left[\left(\frac{g_{ij}}{r}+c_1\right)e^{-\zeta(\frac{g_{ij}}{r
}+c_1)^2}-c_2e^{-\zeta(\frac{g_{ij}}{r})^2}\right],
\end{equation}
where $c_1=1/\sqrt{2\zeta},\ c_2=c_1e^{-1/2}$, $r$ is the average radius of 
cells, $g_{ij}$ is the distance between cell $i$ and cell $j$. The adhesion 
factor $\sigma$ is a measure of the strength of the adhesion. The density of 
desmosomes (typical cell junctions) has a nonuniform distribution across 
different layers indicating varying adhesion strength in different cell 
types~\cite{li2013skinsp,skerrow1989changessp,foty2005differentialsp}. $\sigma$ 
can 
be adjusted for different keratinocytes to reflect this variation. We assume 
that the adhesion strength in proliferating cells is the strongest and 
decreases 
from inner to outer layer in non-proliferating compartment due to decreased 
density of desmosomes.  For example, the corneocytes has reduced adhesion as 
Skerrow et al.~\cite{skerrow1989changessp} reported that the ratio of desmosome 
densities of differentiated layers and corneocyte layer was about 6. For 
calculating the
adhesion forces between different cell types $a$ and $b$, the mean 
$\sigma_{ab}=(\sigma_a+\sigma_b)/2$ is used. We also note that stem cells are 
located on niches and move laterally along the basement membrane.

Parameter $\zeta$ is usually assumed to be 7 with which the adhesive force is 
almost zero when the distance between two cells is larger than the mean cell 
radius $r$. The force increases as the distance increases from zero and reaches 
a max force when the distance is about 0.3$r$ (see Fig.~\ref{fig:adhesion}). As 
the distance increases further, the adhesion declines to zero. For 
computational 
efficiency, the distance between two adjacent cells is approximated by the 
shortest distance between cells represented as rectangles with widths and 
heights identical to the corresponding ellipsoids. The cell-cell distance is 
calculated by
\begin{equation}
g_{ij}=\sqrt{x_{ij}^2+y_{ij}^2}.
\end{equation}
where \begin{eqnarray}
x_{ij}&=&\max\{|x_i-x_j|-\frac{w_i}{2}-\frac{w_j}{2},\,0\}\\
y_{ij}&=&\max\{|y_i-y_j|-\frac{h_i}{2}-\frac{h_j}{2},\,0\}
\end{eqnarray}

\begin{table*}[t]
\centering
\footnotesize\caption{\label{tab:rules}Operation rules for cell events in the 
cell migration model}
\begin{tabular}{ll}
\hline
{\bf Event} & {\bf Operation rule} \\
\hline\hline
$\mathtt{SC}\rightarrow\mathtt{SC+TA}
$& Randomly divide a SC and place daughter cells beside each other$^\dagger$ \\
$\mathtt{SC}\rightarrow\mathtt{SC+SC}$
& Randomly divide a SC and place daughter cells beside each other$^\dagger$  \\
$\mathtt{SC}\rightarrow\mathtt{TA+TA}$
& Randomly divide a SC and place daughter cells beside each other$^\dagger$  \\
$\mathtt{TA}\rightarrow\mathtt{SC}$
& Randomly replace a (normal) TA cell or replace the lowest (psoriatic) TA cell 
by a SC$^\dagger$ \\
$\mathtt{TA}\rightarrow\mathtt{TA+TA}$
& Randomly divide a TA cell and place daughter cells beside each other \\
$\mathtt{TA}\rightarrow\mathtt{GA+TA}$ & Randomly divide a TA cell at the 
boundary and place daughter cells beside each other \\
$\mathtt{TA}\rightarrow\mathtt{GA+GA}$ & Divide the outmost TA cell and place 
daughter cells beside each other \\
$\mathtt{GA}\rightarrow\mathtt{TA}$ & Randomly replace a GA cell at the boundary 
by a TA \\
Differentiation & Replace the outmost cell by a cell of the new type \\
Apoptosis & Randomly choose a cell of proper type and remove \\
Desquamation & Remove the outmost CC \\
\hline
\multicolumn{2}{l}{$^\dagger$SCs and normal TA cells only undergo lateral motion 
along the basement membrane.}
\end{tabular}\label{tab:event}
\end{table*}

The equation of motion for cell migration
\begin{equation}
\mu\frac{d{\bf x}_i}{dt}+{\bf F}_i^r+{\bf F}_i^a=0 \ ,
\end{equation}
\begin{equation}\label{eq:force11}
{\bf F}_i^r=\sum_{j=\mathcal{O}(i)}{\bf F}_{ij}^r \ , \ \ \ {\bf 
F}_i^a=\sum_{j=\Omega(i)}{\bf F}_{ij}^a
\end{equation}
can be numerically solved for individual cells by a finite difference method. 
With a small time step $\Delta t_c$, given the location of cell $i$ at time $t$, 
${\bf x}_i(t)$, one can estimate the cell location ${\bf x}_i(t+\Delta t_c)$ as 
a first-order approximation,
\begin{equation}\label{eq:loc}
{\bf x}_i(t+\Delta t_c)\approx{\bf x}_i(t)-\frac{{\bf F}_i^r(t)+{\bf 
F}_i^a(t)}{\mu}\Delta t_c \ .
\end{equation}\\

\noindent {\em The basement membrane ---} Cell migration is bounded by the 
basement membrane, whose morphology and remodeling must be accounted for. The 
undulant basement membrane at location $x$, is generated by a shifted Gaussian 
function:
\begin{equation}
-H(x)=Ae^{-(x-x_0)^2/\sigma_s^2}-B \ ,
\end{equation}
where parameters $A$, $B$ and $\sigma_s$ model the depth and width of the rete 
ridge. The variable $x_0$ controls the location of a rete ridge along the 
basement membrane. Simulations introduce 10\% fluctuations in parameters $A$ and 
$\sigma_s$ and in cell sizes (width and height) about the means when a new (or 
newly differentiated) cell is produced. Periodical boundary condition was used 
to process cell motion across the left and right boundaries. In the normal 
tissue, proliferating cells including stem cells and TA cells only undergo 
lateral movement along the basement membrane.

The complete algorithm that simulates the stochastic spatiotemporal dynamics of 
the epidermis renewal is outlined as follows.
\begin{enumerate}
\item Simulate the cell population kinetics model with the Gillespie's 
algorithm, and store the time series of cell events.
\item Specify the basement membrane and coordinates of the initial cell 
population and calculate mechanical forces exerted on each individual cells 
using Eqs.~(\ref{eq:ol}), (\ref{eq:af}) and (\ref{eq:force11}), and set the 
simulation time $t=0$ and the time step $\Delta t_c$. 
\item Execute available cell events within [$t$, $t+\Delta t_c$] according to 
the rules described in Table~\ref{tab:event}.
\item Calculate mechanical forces for all affected cells and migrate cells to 
the new coordinates calculated by Eq.~(\ref{eq:loc}), and advance the time 
$t\leftarrow t+\Delta t_c$.
\item Iterate Step 3 until stop.
\end{enumerate}
The model simulation and visualization algorithms are implemented in the Matlab 
and Java Language and are available at 
http://www.picb.ac.cn/stab/epidermal.html.

\subsubsection*{Simulations of psoriatic epidermal homeostasis and treatment by 
UVB irradiation}
Both deterministic and stochastic dynamics of a psoriasis plaque and its 
post-treatment remission after 7-episode UVB irradiations and relapse after 
6-episode UVB irradiation are shown in Fig.~\ref{fig:s3}(a) and (b). The 
dynamic interplays between the cytotoxic effect by the immune system and the 
psoriatic stem cell proliferation are shown in Fig.~\ref{fig:s3}(c) and (d) for 
the 7-episode (effective) and 6-episode (ineffective) treatments, respectively. 
At the pretreatment homeostasis, the cytotoxic effect on PSC (the killing rate) 
balances the net PSC proliferation rate. Upon the start of a UVB irradiation, 
the PSC net proliferation initially increases due to increased 
self-proliferation that tends to fill the SC niche capacity, but however drops 
back during later UVB episodes because the healthy SCs proliferation picks up 
the competition for the niches. In this phase, the cytotoxic rate gradually 
drifts lower because of the reduced PSC population. A successful treatment 
(7-episode in Fig~\ref{fig:s3}(c)) brings the net PSC proliferation below the 
cytotoxicity at the end of the therapy and from there the plaque eventually 
achieves a remission where two rates balance again at the quiescient mode 
homeostasis. By contrast, in a less-dosed regimen (6-episode in 
Fig~\ref{fig:s3}(d)), the net PSC proliferation stays higher than the 
cytotoxicity throughout. Once the treatment stops, PSC proliferation 
progressively outcompetes and healthy SC proliferation for the niches and the 
plaque relapes in the end. 

\begin{figure*}
\centering
\includegraphics[scale=0.11]{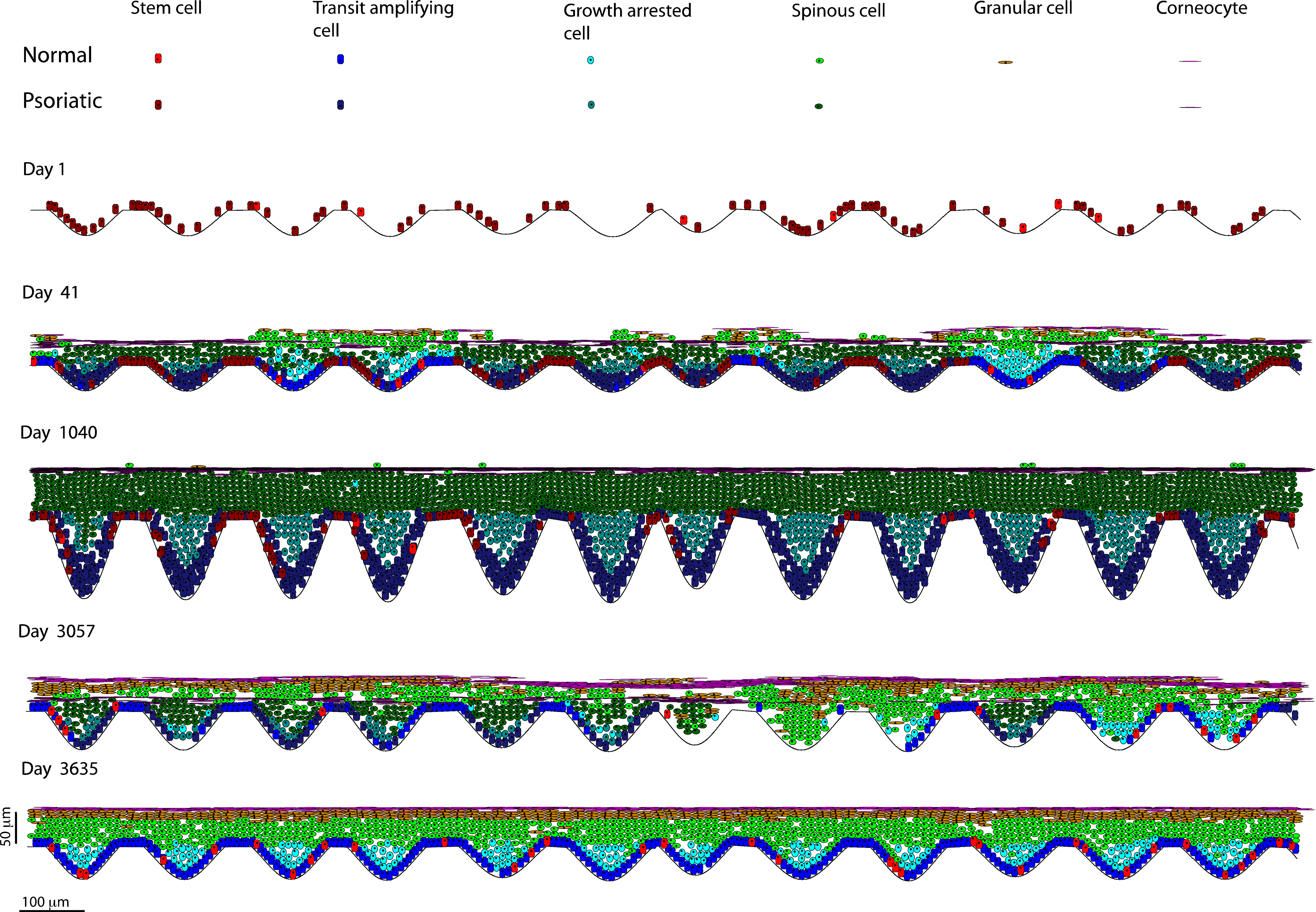}
\caption{\small {\bf Snapshots of simulated psoriatic epidermis under a 
7-episode UVB irradiation (visualization of simulation in Fig. S3(a))}. Each 
snapshot shows two fifth of the simulated length of 5 mm. \label{fig:s4} Same 
types of normal and psoriatic cells are identically colored but with different 
contrasts as annotated in the top panel. The simulation starts at day 1 with a 
mixed initial population of normal (20) and psoriatic (200) stem cells randomly 
distributed on the basement membrane. The simulated tissue grows through a 
transient phase (Day 41, a mix of normal and psoriatic cells are visible. Note 
that in this phase the disparity in growth rates of normal and psoriatic 
keratinocytes generate morphological heterogeneity in the epidermis.) to 
achieve 
a parallel homeostasis between normal and psoriatic keratinocytes (Day 1040, 
most cells are psoriatic and the epidermis thickens with extended rete ridges). 
Episodes of UVB irradiation reduce the cell population (Day 3057, at the early 
recovery with normal cells). After recovery, the epidermis is composed 
primarily of normal keratinocytes (Day 3635).  In simulation, the rete ridge in 
psoriasis is dynamically extended (see the text for the dynamic model of rete 
ridge) to reach a height of 126 $\mu$m at the homeostasis (about 3-4 times 
larger than that of the normal 
tissue~\cite{iizuka2004psoriaticsp,weatherhead2011keratinocytesp}). The 
simulation sets a small $\Delta t_c=0.002$ day to ensure accurate numerical 
integration and to bracket within each migration time step only a few (about 2) 
cellular events happen along a unit epidermal length (1 mm). Psoriatic TA cells 
are not anchored but migrate within a confined region above the basement 
membrane with a height depending on the TA cell density.} 
\end{figure*}

Fig.~\ref{fig:s4} shows a few snapshots of 2D tissue stratification during a 
simulation. Simulation movies are available at URL: 
http://www.picb.ac.cn/stab/epidermal.html. 

In the psoriatic tissue, because of high density of proliferating cells, TA 
cells are allowed to migrate within a larger (but restricted) area, within a 
band parallel to the basement membrane. The height of the band is modulated by 
the density of proliferating cells with a 15 $\mu$m increment (decrement) for 
every 18000 mm$^{-2}$ increase (decrease) in proliferating cell density.

In comparison to a fixed rete ridge, the dynamics of rete ridge remodeling is 
considered in the psoriasis model. We assume that the rete ridge remodels 
between normal and psoriatic structures according to proportions of normal and 
psoriatic cells. The fraction of normal progenitor cells is defined as
\begin{equation}
 {r=\tt\frac{normal\  SC+TA\  cells}{total\  SC+TA\  cells}} \ ,
\end{equation}
Let $y$ be the rete ridge height in time and $Y$ is the target homeostatic rete 
ridge height. We assume that the dynamics of the rete ridge follows a 
first-order dynamics
\begin{equation}\label{eq:reteode}
\tau\frac{dy}{dt}=Y-y \ ,
\end{equation}
where $\tau$ is the time constant for the dynamics. We further assume $Y$ is 
linearly regulated by $r$:
\begin{equation}\label{eq:linear}
Y=Y_{\rm max}-(Y_{\rm max}-Y_{\rm min})r \ ,
\end{equation}
where $Y_{\rm max}$ ($Y_{\rm min}$) is reached at $r=0$ ($r=1$).

Eq.~(\ref{eq:reteode}) is numerically solved by the finite difference method as 
follows:
\begin{equation}
 {y(t+\Delta t_{\rm bm}) = \frac{Y(t)-y(t)}{\tau}\Delta t_{\rm bm} + y(t)}, 
\end{equation}
where $Y(t)$ is calculated from Eq.~(\ref{eq:linear}).

We set $Y_{\rm min}=$40 $\mu$m to match the healthy tissue when normal stem 
cells dominate at the homestasis and $Y_{\rm max}=$126 $\mu$m, about 3 times 
higher than the normal epidermis. Based on data from successful UVB treatments 
showing that psoriasis plaques achieved remission in about 3-4 
months~\cite{Dawe1998sp,trehan2002sp,Hartman2002sp}, we set the time constant 
$\tau=$100 day and $\Delta t_{\rm bm}$ was chosen as 20 day. In a simulation, 
whenever the basement membrane remodels, all stem cells and normal TA cells are 
relocated along with the membrane accordingly and other cells stay above the 
basement membrane.

\subsubsection*{Identification of model parameters}
We archive parameter values in Table 2 and 3 in the main text. The literature 
that reported parameter values or provided information to derive the parameter 
values is listed in the tables. For example, cell sizes of keratinocytes at 
different stages are reported in histology studies, and rate constants for 
proliferation and differentiation are assigned based on measurements (such as 
the stem cell proliferation rate) or identified from cell density distribution 
data and the measured epidermal turnover time. We list multiple references for 
a 
parameter value if available for confirmation, but we acknowledge that the 
reference list may not be exhaustive. We also note that these parameter values 
serve as references and should not be considered as fixed because of variations 
of the epidermis homeostasis at skin location, age, gender and pathological 
conditions and uncertainties in experimental measurements. 

Parameters that are not measured are estimated or assumed (as indicated in the 
Tables). Assumed parameter values are justifiable based on available 
information. For example, the stem cell growth capacity $p_{sc}^{\rm max}$ in 
the healthy tissue is assumed as 4.5e3 mm$^2$, about 2 times the homeostatic 
density, considering significant loss of stem cells through symmetric division 
and apoptosis. In comparison, the growth capacity in the psoriatic tissue is 
extended ($\lambda=3.5$ fold larger), which was suggested by the histology data 
that showed extending rete ridge. Correspondingly, the maximum rete ridge 
height 
$Y_{\rm max}$ is assumed more than 3 times that of the healthy one. Parameters 
such as the maximum cytotoxic rate $K_p$ have not been experimentally measured, 
which is chosen to predict a psoriasis mode (e.g., $K_p=6$ mm$^{-2}$day$^{-1}$ 
is chosen to parameterize a switchable psoriatic tissue).  

%\bibliographystyle{unsrt}
%\bibliography{reference}

\end{document}